\documentclass[a4paper,fleqn,usenatbib]{mnras}

\usepackage{newtxtext,newtxmath}

\usepackage[T1]{fontenc}
\usepackage{ae,aecompl}

\usepackage{graphicx}	
\usepackage{amsmath}	
\usepackage{amssymb}	
\usepackage{pdflscape}  
\usepackage{subfig}     
\usepackage{xcolor}

\newcommand{\ms}{M_{\odot}}
\newcommand{\ns}{M\!N\!S}
\newcommand{\ams}{\alpha_{M\!N\!S}}

\title[A new DTD for MNS]{A new delay time distribution for merging neutron stars tested against Galactic and cosmic data}

\author[P. Simonetti et al.]{
Paolo Simonetti,$^{1,2}$\thanks{E-mail: paolo.simonetti@inaf.it}
Francesca Matteucci,$^{1,2,3}$
Laura Greggio,$^{4}$
Gabriele Cescutti$^{2}$
\\
$^{1}$Dipartimento di Fisica, Sezione di Astronomia, Universit\`a degli Studi di Trieste, Via G. B. Tiepolo 11, I-34143 Trieste, Italy\\
$^{2}$INAF, Osservatorio Astronomico di Trieste, Via G.B. Tiepolo 11, I-34143 Trieste, Italy\\
$^{3}$INFN, Trieste, Via A. Valerio 2, I-34127 Trieste, Italy\\
$^{4}$INAF, Osservatorio Astronomico di Padova, Vincolo dell'Osservatorio 5, I-35122 Padova, Italy
}

\date{Submitted to MNRAS}

\pubyear{2019}

\begin{document}
\label{firstpage}
\pagerange{\pageref{firstpage}--\pageref{lastpage}}
\maketitle

\begin{abstract}
The merging of two neutron stars (MNS) is thought to be the source of short gamma-ray bursts (SGRB) and gravitational wave transients, as well as the main production site of r-process elements like Eu. We have derived a new delay time distribution (DTD) for MNS systems from theoretical considerations and we have tested it against (i) the SGRB redshift distribution and (ii) the Galactic evolution of Eu and Fe, in particular the [Eu/Fe] vs [Fe/H] relation. For comparison, we also tested other DTDs, as proposed in the literature.
To address the first item, we have convolved the DTD with the cosmic star formation rate, while for the second we have employed a detailed chemical evolution model of the Milky Way. We have also varied the DTD of Type Ia SNe (the main Fe producers), the contribution to Eu production from core-collapse SNe, as well as explored the effect of a dependence on the metallicity of the occurrence probability of MNS. 
Our main results can be summarized as follows: (i) the SGRB redshift distribution can be fitted using DTDs for MNS that produce average timescales of 300-500 Myr; (ii) if the MNS are the sole producers of the Galactic Eu and the occurrence probability of MNS is constant the Eu production timescale must be on the order of $\lesssim$ 30 Myr; (iii) allowing for the Eu production in core-collapse SNe, or adopting a metallicity-dependent occurrence probability, allow us to reproduce both observational constraints, but many uncertainties are still present in both assumptions.
\end{abstract}





\section{Introduction}

The merging of two neutron stars via the emission of gravitational waves has been invoked to explain a variety of phenomena like the production of heavy r-process elements \citep{korob12,hoto13} and the short gamma-ray bursts (SGRBs) \citep{eichler89,narayan92}. The observation of binary systems composed by compact and massive star remnants \citep{hulse75,tauris17} with measurable changes in their orbital periods, and the recent detection of the gravitational wave (GW) transient GW170817 \citep{abbott17a} strongly support the occurrence of this process.

Elements heavier than Fe cannot be produced by exoenergetic fusion reactions in stars and instead they must be the result of neutron capture on Fe-peak nuclei. If the ingestion of neutrons is "rapid" compared to the $\beta$ decay timescales, heavy nuclei are produced,  while "slow" captures result in relatively lighter products \citep{burbidge57,seeger65}. Correspondingly, these elements are called r- and s-process elements. Moreover, r-process elements can be subdivided in a light (with A $<$ 90) and a heavy (A $>$ 90) subclass. A widespread used tracker for heavy r-process elements is Europium (Eu) \citep[e.g.][]{matteucci14}. The production of heavy r-process elements requires very neutron rich environments, such as core-collapse SNe (CC-SNe) and MNS. The dominant production site, though, remains uncertain \citep{cote18b}.

Early studies, based on the observation of r-process elements in very metal poor ([Fe/H] $<$ -3.0 dex) stars, led to the conclusion that r-process production should occur in massive, short-lived stars (m $>$ 10 $\ms$) \citep{truran81}. 

Hydrodynamical simulations \citep[e.g.][]{woosley94,wheeler98} supported such a result, showing that the neutrino-driven wind from newly born neutron stars (NS) in stars with masses higher than 20 $\ms$ is a promising (although imperfect) site for r-process nucleosynthesis. However, subsequent studies \citep{arcones07,arcones11,mp12} have cast doubts on the capability of this mechanism to produce a robust r-process abundance pattern. In particular, it was found that neutrino-driven winds were not enough neutron rich to produce elements with A $>$ 120. Moreover, the final abundance pattern is very sensitive to details of the physical conditions at explosion. A rare sub-class of metal poor CC-SNe, called magneto-rotational (MR) SNe, has been theorized to be able to overcome these problems \citep{cameron03,winteler12}, though their ability to reproduce the solar abundances of these elements is not clear \citep{woosley06}, but see also \citep{cescutti14}.

In parallel, other authors have explored the role of compact object mergers (CBM) (like NS-NS and NS-BH systems) in the production of r-process elements. Nucleosynthesis calculations \citep{korob12,eichler15} predict the ejection of up to $10^{-2}\, \ms$ of r-process matter with a robust abundance pattern in a single coalescence event, due to the large number of neutrons per target Fe nuclei. Therefore, this source appears to be more reliable with respect to the CC-SNe.

The viability of double neutron star mergers as r-process production site has been tested in galactic chemical evolution simulations, which trace the abundances of different elements in the gas of the Milky Way. In this regard, a very important parameter to evaluate is the timescale of production of a given element, which can be estimated as the average coalescence timescale of a single burst stellar population. For MNS systems, the delay time (or time between the birth of the binary system and its final merging) is determined by both the stellar nuclear lifetime and the gravitational delay time, and it can vary in a wide range. The typical timescale for enrichment from MNS depends on the distribution of the delay times (DTD). 

A popular choice in literature consist in selecting a DTD in the form of a power law, i.e. $\propto t^{\gamma}$ with $0.5 \le\gamma\le 2.0$ \citep{cote17a,hoto18} where $t$ is the total (i.e. nuclear+gravitational) coalescence delay time; in other studies, the total coalescence delay time is obtained by adding a constant gravitational delay on top of stellar evolutionary lifetimes \citep{matteucci14,vangioni15}. Some authors, like \citet{matteucci14}, have tested both MNS only, and a combination of MNS and CC-SNe as Eu producers. In all studies it is found that, in order to reproduce the decreasing trend in the [Eu/Fe] vs [Fe/H] relation in the thin disk stars, the Europium pollution timescale should not exceed 100 Myr or be even shorter (10-30 Myr).These timescales can be attained either using very steep power laws ($\propto t^{-2}$) or assuming a constant gravitational delay of 1 Myr on top of the nuclear evolution timescale of the progenitors. These DTDs are in disagreement with the results of population synthesis models \citep{belc17,giacobbo18b,cote18b} that predict a $\gamma$ in the range $[-1.0;-1.5]$. 

Gamma-ray bursts (GRBs) are luminous transients of high energy photons (in the range of 10-100 keV) incoming isotropically to the Earth at a rate of 1 per day \citep{klebe73}. They can be divided in short ($<2$ s, SGRBs) and long ($>2$ s, LGRBs) \citep{norris84,meszaros03}; SGRBs are characterized by high peak energy, while LGRBs exhibit a softer emission. It is generally accepted that short and long GRBs are produced by different mechanisms; in particular the latter have been explained as due to the acceleration of matter along the poles of a newly born black hole during massive CC-SNe explosions \citep{woosley99}. The SGRBs are instead thought to be produced from the magnetic recombination or neutrino-antineutrino annihilation during the merging of compact binary objects \citep{narayan92}.

Despite the small statistics, various groups \citep{ghirlanda16,zhang18} have reconstructed the redshift distribution of SGRBs, while others \citep{davanzo15,fong17} have tried to reproduce it by convolving different DTDs with the cosmic star formation history. It has been found that the SGRB redshift distribution is better reproduced when a DTD $\propto t^{-1}$ or $\propto t^{-1.5}$ is used. Moreover, between 1/4 and 1/3 of the total SGRBs comes from early-type galaxies \citep{berger14,fong17}, which host old stellar populations. This evidence contrasts with the notion of short timescales to describe MNS, in favor of longer ones.

A great deal of information comes from the detection of GW170817. Not only it has been the first, direct evidence of a MNS \citep{abbott17a}, but it was also associated with a SGRB \citep{abbott17b} and an optical transient known as "kilonova" \citep{smartt17}. The study of its multiband lightcurves and of the spectra has provided constraints on the composition of the ejecta, though a consensus has not yet been reached. \citet{smartt17} concluded that these ejecta should have been rich in light r-process and very deficient in heavy r-process, while others \citep{tanvir17,troja17,evans17,pian17} support the hypothesis of a robust production of extremely opaque r-process elements. Finally, based on this event, \cite{abbott17a} estimated the cosmic local rate of MNS events to be $1540_{-1220}^{+3200}$ $\text{yr}^{-1}\text{Gpc}^{-3}$.

In this paper we want to test under which conditions it is possible to accomodate the various observational constraints under the hypothesis that MNS are responsible for all SGRBs, and that they are important contributors to the Europium chemical enrichment of the Milky Way. To do this we: (i) derive a new formulation for the DTD from theoretical considerations, adapting the \citet{greggio05} procedure, which was developed for Type Ia Supernovae, to the progenitors of MNS; (ii) use the cosmic star formation history (CSFR) and the redshift distribution of SGRBs to evaluate the fraction of massive stars which give rise to a MNS event; (iii) test the results versus the Milky Way abundance pattern of Eu and Fe using a full galactic chemical evolution model of the Milky Way, which successfully reproduces a majority of observational constraints \citep[e.g.][]{matteucci09}. We also explore the effect of adopting the DTD formulations which are proposed in the literature.

\section{Our new DTD}

The delay time distribution function $f_{M\!N\!S}(\tau)$ for MNS represents the distribution of the coalescence times from an instantaneous burst of star formation of unitary mass and it is fundamental to compute the MNS rate. The MNS rate can be written as:
\begin{equation}
\label{cap3.snrate}
R_{\ns}(t)=k_{\alpha}\int_{\tau_i}^{\text{min}(t,\tau_x)} \alpha_{\ns}(\tau)\, \psi(t-\tau)\,f_{\ns}(\tau)\, d\tau 
\end{equation}
where $\psi$ is the SFR, $k_{\alpha}$ is the number of stars with mass in a suitable range per unit mass of star forming gas in a stellar generation, and $\alpha_{\ns}$ is the fraction of them which gives rise to MNS events. In principle, both $k_\alpha$ and $\alpha_{\ns}$ can vary as a function of time, but for the sake of simplicity we will assume them constant. In Sect. 5.4 we will explore the effect of varying $\alpha_{\ns}$. The time $\tau$ is the delay time defined in the range $(\tau_i,\tau_x)$, where $\tau_i$ is the minimum delay time of occurrence for merging neutron star (here fixed to 10 Myr) and $\tau_x$ is the maximum delay time which can be larger than a Hubble time, depending on the progenitor model. Since $f_{\ns}(\tau)$ is a distribution function, it must be normalized to 1 in the allowed range for $\tau$:
\begin{equation}
\int_{\tau_i}^{\tau_x} f_{\ns}(\tau)\, d\tau = 1.
\end{equation}
The parameter $k_\alpha$ depends on the IMF, and it is:
\begin{equation}
k_{\alpha}=\int_{m_L}^{m_U} \phi(m) \, dm
\end{equation}
where $m_L$ and $m_U$ are, respectively, the progenitor minimum and maximum mass to produce a NS. In this paper we adopt $m_L = 9 \ms$ and $m_U$ = 50 $\ms$.

We now derive the form of $f_{\ns}(\tau)$. For MNS the time delay is the sum of the evolutionary lifetime of the secondary component of the binary system plus the gravitational time delay. From \citet{landau62}:

\begin{equation}
\tau_{gw} = \frac{0.15A^4}{m_1m_2(m_1+m_2)} \, Gyr
\label{cap3.land}
\end{equation}

where $m_1$ and $m_2$ are the masses of the primary (more massive) and of the secondary star of the binary system, respectively, while $A$ is the initial separation of the neutron star binary system. 
We can now show that this equation can be rewritten as a function of the total mass, instead of the masses of the two components,  with a small error. The mass dependent term, namely the denominator of \eqref{cap3.land}, can be written as:
\begin{equation}
\mu(M_{D\!N},m_2)=M_{D\!N}^2\cdot m_2 - m_2^2 \cdot M_{D\!N},
\end{equation}
with $M_{D\!N}=m_1+m_2$ i.e. the total mass of the system. This function can be studied in its two variables. For a fixed $M_{D\!N}$ this formula produces a parabola with the maximum at $m_2=0.5M_{D\!N}$.

\begin{figure}
\includegraphics[width=\columnwidth]{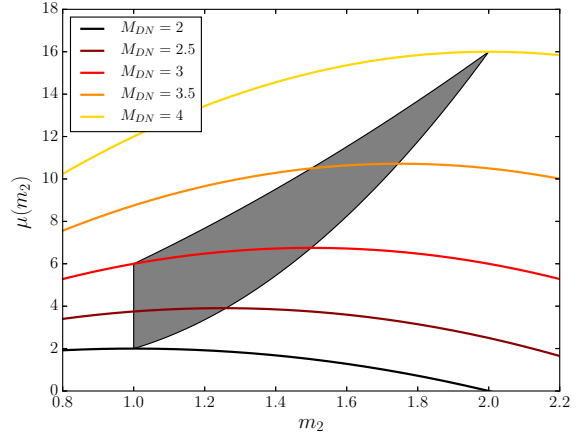}
\caption{The function $\mu(M_{D\!N},m_2)$ is plotted with respect to $m_2$ for five different fixed values of $M_{D\!N}$ (2.0 $\ms$ in black, 2.5 $\ms$ in dark red, 3.0 $\ms$ in red, 3.5 $\ms$ in orange and 4.0 $\ms$ in yellow solid lines). The gray shaded area represents the space of valid values for $m_2$ (i.e. represents a system with a secondary lighter than the primary and in the correct mass range, for a fixed total mass of the system).}
\label{sec2.approx}
\end{figure}

The lightest known neutron star whose mass has been precisely measured is the companion of the binary pulsar PSR J0453+1559, with an estimated mass of $1.17 \pm 0.004$ $\ms$ \citep{martinez15}. On the other hand, the heaviest measured neutron star is the PSR J0348+0432, with a mass of $2.01 \pm 0.04$ $\ms$ \citep{antoniadis13}. In a yet not fully confirmed study, \citet{linares18} estimate the mass of PSR J2215+5135 to be $2.27 \pm 0.15$ $\ms$, and this fact would rule out some of the proposed equations of state for NS interior \citep{ozel16}.

We have decided to adopt a slightly different range with a lower limit to the mass of the neutron star of 1 $\ms$ and an upper limit of 2 $\ms$. Therefore the allowed range for $M_{D\!N}$ goes from 2.0 to 4.0 $\ms$. The requirements on the mass of the primary NS further constrains the range of possible values for $m_2$ so that not all combinations of $M_{D\!N}$ and $m_2$ are acceptable. In figure \ref{sec2.approx} we show the function $\mu(M_{D\!N},m_2)$ plotted vs $m_2$ for five different values of $M_{D\!N}$. The shaded are shows the portion of the plane in which $m_1 > m_2$. The sections of the parabolae within the shaded area are very flat: for each $M_{D\!N}$ the function $\mu$ varies very little (no more than 10\%) inside the allowed region of the parameter space ($m_1$,$m_2$).

We can conclude that it is possible to rewrite $\mu$ as a function of $M_{D\!N}$ with a negligible error: every binary system with the same total mass and the same initial separation has approximately the same delay. Substituting the value of the maximum for each $M_{D\!N}$, $\mu(M_{D\!N},m_2)$ becomes:
\begin{equation}
\label{cap3.disegno}
\mu(M_{D\!N})=0.25M_{D\!N}^3
\end{equation}
and so it follows that:
\begin{equation}
\label{cap3.greg1}
\tau_{gw} = \frac{0.6A^4}{M_{D\!N}^3} Gyr.
\end{equation}
Having simplified the Landau formula we derive distribution of the delay times due to the emission of gravitational waves as follows.

The contribution to the number of systems with delay $\tau_{gw}$ from systems with separation $A$ and total mass $M_{D\!N}$ can be written as:
\begin{equation}
\label{cap3.dn}
df(\tau_{gw})=df(A,M_{D\!N})=g(A)\,h(M_{D\!N})\,dA\,dM_{D\!N},
\end{equation}
where $g(A)$ is a function describing the initial separations, while $h(M_{D\!N})$ is the distribution of the total masses of the systems that will merge. In Eq. (\ref{cap3.dn}) these two functions are assumed to be independent. We further assume that $g(A)$ can be described as a power law and, for simplicity, that the distribution of binary masses is flat. Therefore the two functions can be written as:
\begin{equation}
\begin{cases}
g(A) \propto A^{\beta} \\
h(M_{D\!N}) \propto const.
\end{cases}
\label{cap3.gh}
\end{equation}
As can be seen, $\beta$ parametrizes the shape of the distribution in initial separations. Integrating Eq. \eqref{cap3.dn} over the relevant range of separations we derive the number of MNS systems which merge with a gravitational time delay $\tau_{gw}$:
\begin{equation}
\label{cap2.int}
f(\tau_{gw})d\tau_{gw}=d\tau_{gw}\int_{A_l}^{A_s} g(A)\,h(M_{D\!N})\,\Bigl |\frac{\partial M_{D\!N}}{\partial \tau_{gw}}\Bigr | \,dA
\end{equation}
with $A_l$ and $A_s$ are the minimum and maximum separations that contribute to $n(\tau_{g\!w})$ for systems with total mass $M_{D\!N}$. They can be calculated as:
\begin{equation}
\label{cap2.A}
A_x=\Bigl (\frac{M_{D\!N,x}^3\tau_{gw}}{0.6}\Bigr )^{1/4}
\end{equation}
where the subscript x stands for s and l. The result of such calculation is:
\begin{equation}
\label{cap2.A1}
f(\tau_{gw}) \propto \frac{1}{\tau_{gw}^{4/3}} \frac{1}{\beta+7/3}[A^{\beta+7/3}]_{A_l}^{A_s}.
\end{equation}
Substituting Eq. \eqref{cap2.A} in Eq. \eqref{cap2.A1} we derive the distribution of the gravitational wave delay times as:
\begin{equation}
f(\tau_{gw}) \propto \tau_{gw}^{(1/4)\beta-3/4}(M_{D\!N,s}^{(3/4)(\beta+7/3)}-M_{D\!N,l}^{(3/4)(\beta+7/3)}).
\end{equation}

\begin{figure}
\includegraphics[width=\columnwidth]{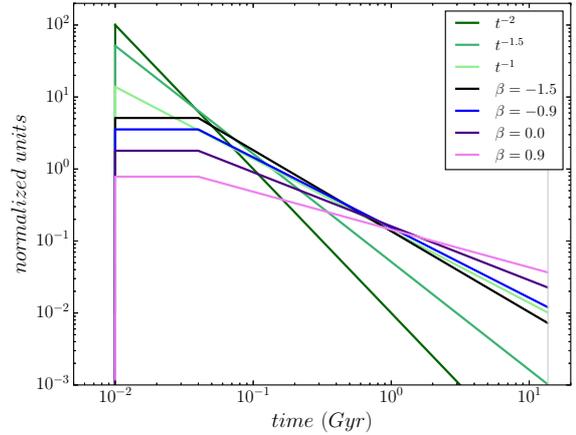}
\caption{The seven different DTDs tested for MNS. For the DTD derived in this paper, the three phases are clearly visible: the initial null plateau, the plateau representing the close binaries that promptly merges and the tail for wide binary systems. We have tested it for four different values of $\beta$: -1.5 (black), -0.9 (blue), 0.0 (indigo) and 0.9 (purple). The three DTDs $\propto t^{-\gamma}$ with $\gamma$ equal to -1 (light green), -1.5 (medium green) and -2 (dark green) are also shown. The area under each of these curves is the same and equal to 1.}
\label{sec2.DTD}
\end{figure}

The number of systems with a total delay between $\tau$ and $\tau+d\tau$ will be close to the number of systems with a gravitational delay between $\tau_{gw}$ and $\tau_{gw}+d\tau_{gw}$, because of the short and similar nuclear lifetime ($\tau_{n}$) of stars with high masses, that ranges between $\sim 30$ Myr for 9 $\ms$ stars and $\sim 4$ Myr for 50 $\ms$ stars. However, the existence of a distribution of $\tau_n$ implies that the DTD deviates from $f(\tau_{gw})$ at the short delay times, because the parameter space is limited from the requirement $\tau =\tau_n + \tau_{gw}$. \citet{greggio05} developed a DTD for binary white dwarfs taking this effect into account. Following her results we construct the following distribution of total (nuclear + gravitational) delay time:
\begin{equation}
\label{cap2.final}
f(\tau) \propto
\begin{cases}
0 \,\,\,\quad  \text{if} \quad \tau<10 \, Myr \\
p_1 \quad  \text{if} \quad 10<\tau<40 \, Myr\\
p_2 \, \tau^{0.25\beta-0.75}(M_{D\!N,s}^{0.75(\beta+2.33)}-M_{D\!N,l}^{0.75(\beta+2.33)})\\
\quad \quad \text{if} \quad 40\, Myr<\tau<13.7 \, Gyr
\end{cases}
\end{equation}
where $p_1$ and $p_2$ will be chosen so as to obtain a continuous and normalized function. The first portion of the distribution ends with the formation of the first double neutron star system. Ten million years  is the nuclear lifetime of a typical massive star. The second portion refers to systems which merge soon after the formation of the double neutron star binary system. Similar to the distributions in Greggio (2005), this portion of the distribution function is described as a flat plateau, up to the maximum nuclear time delay of the double neutron star binary or, in other words, the lifetime of the minimum mass progenitor of a neutron star ($\sim 9$ $\ms$). The third portion is just the distribution of the gravitational delay times and attains to those systems for which the time delay is dominated by the gravitational radiation mechanism described by the Landau equation. The results can be seen in figure \ref{sec2.DTD} where a normalization factor of 1 has been assumed.

\subsection{Other tested DTDs}

As a comparison for DTDs, we have also tested four other common options found in literature. Three of them are single-slope power laws, with $\gamma=-1.0,\,-1.5,\,-2.0$, as proposed by \citet{cote18a,cote18b}. We again impose a minimum value to the total delay time of 10 Myr, which represents the shortest lifetime of neutron stars progenitors. All of them are normalized between this effective minimum time and one Hubble time (13.7 Gyr). The various choices of DTDs discussed so far are shown in figure \ref{sec2.DTD}.

We also tested the DTD implemented in \citet{matteucci14}, in which the total delay time is equal to 1 Myr on top of the stellar lifetime, implicitely assuming that all binary neutron stars are born very close, and merge soon after they are formed. In \cite{matteucci14} it has been shown that longer gravitational delays do not reproduce the Eu abundances in very metal poor stars.

\begin{figure}
\includegraphics[width=\columnwidth]{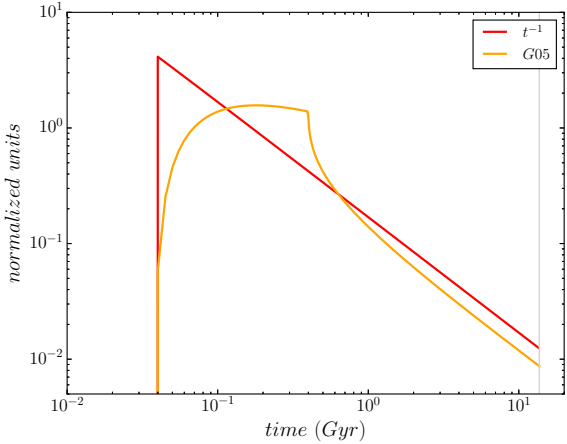}
\caption{The two different DTDs used to represent the SNeIa in this paper: the $\propto t^{-1}$ suggested by \citet{totani08} and \citet{maoz12} (red line) and the DTD described in \citet{greggio05} in the wide double degenerate case with $\beta_a=-0.9$ (orange line). This value for $\beta_a$ is also suggested by \citet{matteucci09}. The area under each of these curves is the same and equal to 1, i.e. they are normalized.}
\label{sec2.SNeIa}
\end{figure}

\begin{figure}
\includegraphics[width=\columnwidth]{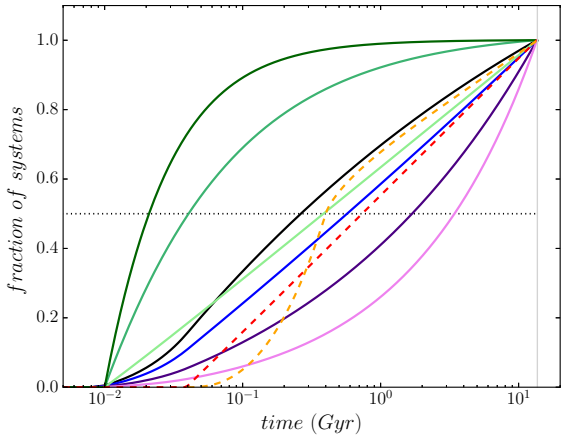}
\caption{The cumulative distributions associated with the DTDs tested for MNS (solid lines) and for SNeIa (dashed lines). The color code is the same as used in figures \ref{sec2.DTD} and \ref{sec2.SNeIa}. The dotted line marks the 0.5 value for fraction of merged binary neutron stars and provide an immediate comparison between the different timescales produced by the DTDs.}
\label{sec2.Cumulative}
\end{figure}

\subsection{The progenitors of SNeIa}

Since one of our constraints concerns the trend of [Eu/Fe] vs [Fe/H] and since an important part of the iron enrichment comes from SNeIa, we need to specify a choice for their DTD. We test two different options:

\begin{enumerate}[i.]
\item a distribution proportional to $t^{-1}$, as suggested by \citet{totani08} and \citet{maoz12} on the basis of empirical data;
\item the DTD in the wide double degenerate scenario depicted in \citet{greggio05} with an initial separation distribution coefficient $\beta_a=-0.9$ and a maximum nuclear lifetime $\tau_{n,x}=0.4$ Gyr. The reasons for these choices are presented in \citet{matteucci09}.
\end{enumerate}

Both DTDs are normalized in the range 0.04-13.7 Gyr, and we show them in figure \ref{sec2.SNeIa}. These two distributions are not very different. In fact, at times larger than $\sim 3$ Gyr, the fraction of exploded SNeIa is similar. On the other hand, as will be show in section 5, their effect on the [Eu/Fe] vs [Fe/H] relation at low (-2.5 < [Fe/H] < -1.0 dex) metallicities is pronounced. Chemical evolution models based on DTDs very different from this kind of shape have difficulties to fit the abundances of low metallicity stars.

The various DTDs discussed so far imply different timescales for the Europium and Iron pollution on the ISM. As mentioned in the Introduction, the pollution timescale can be represented as the average coalescence timescale for a single burst stellar population, or the time within which half of the events occur. Figure \ref{sec2.Cumulative} shows the cumulative distribution of the DTDs for the MNS and the SNeIa's adopted in our computations. The Europium pollution timescales for the various models are very different, ranging from $\sim 20$ Myr to 2 Gyr as the DTD for the MNS changes from the steepest pure power law to the flattest of our proposed DTD. Instead, in our models, the timescale for Fe production from SNeIa ranges only from 0.4 to 0.7 Gyr.

\section{Comparison with the SGRB redshift distribution}

As briefly described in the Introduction, it has been possible to derive the SGRB redshift distribution despite the small statistics. However, there are discrepancies between the results of different groups.

\subsection{The SGRB redshift distribution}

To constrain our models we consider two redshift distributions of SGRB proposed in the literature: the \cite{ghirlanda16} (herefater G16) and \cite{zhang18} (hereafter ZW18).

G16 considered seven observables: (i) peak flux, (ii) fluence, (iii) observer frame duration, (iv) observer frame peak energy distribution, (v) redshift, (vi) isotropic energy and (vii) isotropic luminosity, usually not available at the same time for each event, and a set of parametric relations that bind these together. The parametric form chosen for the redshift is:
\begin{equation}
\Psi(z)=\frac{1+p_1z}{1+(z/z_p)^{p_2}}
\end{equation}
Then, they generated synthetic populations of SGRBs, choosing randomly the values of the parameters, and compared their features to the observational dataset. A total of 10 free parameters (3 for redshift distribution, 3 for peak energy distribution and 4 for energy-luminosity correlations) have been tested. This is referred in the paper as 'model a'. G16 considered also a slightly different scheme for relations between observables with independent (from the peak energy and between themselves) distributions for luminosity and duration, raising the number of free parameters to 11. This is referred as 'model c'.

The values of the redshift distribution parameters which best represent the observational characteristics  of the SGRB population turn out to be: $p_1$ = 2.8 (model a) or 3.1 (model c), $p_2$ = 3.5 (model a) or 3.6 (model c) and $z_p$ = 2.3 (model a) or 2.5 (model c). 

With a different approach, ZW18 first considered the 16 SGRBs with known redshift. From them, they derived a peak energy - luminosity relation and have used this relation to find the redshift of the other 284 SGRBs in the \textit{Fermi-GBM} sample. Finally, ZW18 have used these inferred redshifts to fit a bimodal relation in the form of:
\begin{equation}
\Psi(z)\propto
\begin{cases}
(1+z)^{-3.08} \quad z< 1.60 \\
(1+z)^{-4.98} \quad z\ge 1.60
\end{cases}
\end{equation}
This is an ever-decreasing function of redshift, sharply in contrast with the result of G16.

\subsection{Simulations and results}

In this section we compare the SGRB redshift distribution shown above to theoretical expectations of the redshift distribution of MNS based on the DTDs described in Sect. 2.  To do this we convolve the DTDs with the cosmic star formation rate (CSFR) (see Eq. \eqref{cap3.snrate}), for which we adopt the relation obtained by \citet{madau14}:
\begin{equation}
\Psi(z)=\frac{0.015(1+z)^{2.7}}{1+((1+z)/2.9)^{5.6}}\ \ \   \ms \text{Mpc}^{-3} \text{yr}^{-1}
\end{equation} 
base on a \citet{salpeter55} IMF, that is:
\begin{equation}
\label{cap2.sal}
\phi(m)=0.171 \  m^{-2.35},
\end{equation}
when normalized between 0.1 and 100 $\ms$.

\begin{table}
\begin{center}
\begin{tabular}{lccc}
\hline
MNS & $\ams$ & merged & merged \\
DTD & ($\times 10^{-2}$) & in 20 Myr & in 100 Myr \\
\hline
Our DTD $\beta=0.9$ & $0.66$ & 0.008 & 0.056\\
Our DTD $\beta=0.0$ & $0.81$ & 0.019 & 0.129\\
Our DTD $\beta=-0.9$ & $1.02$ & 0.039 & 0.241\\
Our DTD $\beta=-1.5$ & $1.18$ & 0.056 & 0.336\\
\hline
Prop to $t^{-1}$ & $1.09$ & 0.096 & 0.319\\
Prop to $t^{-1.5}$ & $1.60$ & 0.301 & 0.703\\
Prop to $t^{-2}$ & $1.73$ & 0.500 & 0.901\\
Constant 10 Myr &  $1.75$ & 1.000 & 1.000\\
\hline
\end{tabular}
\end{center}
\caption{The table reports the value of the occurrence probability $\ams$ for each tested DTD that must be inserted in order to obtain the present time cosmic rate of MNS at $z=0$ equal to the gravitational wave event rate found by the LIGO/Virgo Collaboration \citep{abbott17a}. In the third and fourth columns are reported the fractions of MNS systems merged after 20 Myr and 100 Myr from a single initial starburst.}
\label{cap5.ams}
\end{table}
 
For a Salpeter IMF and a 9-50 $\ms$ range for progenitors of NS we have  $k_{\alpha}$ = 0.0059 neutron star progenitors per $\ms$ of star-forming gas. The parameter $\ams$ has been chosen in order to reproduce the current rate of GW events per unitary volume of Universe as found by Abbott et al. (2017a), $1540$ $ yr^{-1} Gpc^{-3}$. The values for $\ams$ are reported in table \ref{cap5.ams} for the different DTDs, alongside with the fractions of MNS systems merged after 20 Myr (third column) and 100 Myr (fourth column) after an initial single star formation burst. Together with figure \ref{sec2.Cumulative} this allows us to make a simple comparison among the timescales of the different distributions. 

In this way we have computed a redshift distribution of MNS rate directly comparable with the derived redshift distributions of SGRBs of G16 and ZW18. Both their relations have been multiplied by a suitable factor to reproduce the current GW rate, in order to make a comparison possible.

\begin{figure*}
\centering
\subfloat[]{\includegraphics[width=0.5\textwidth]{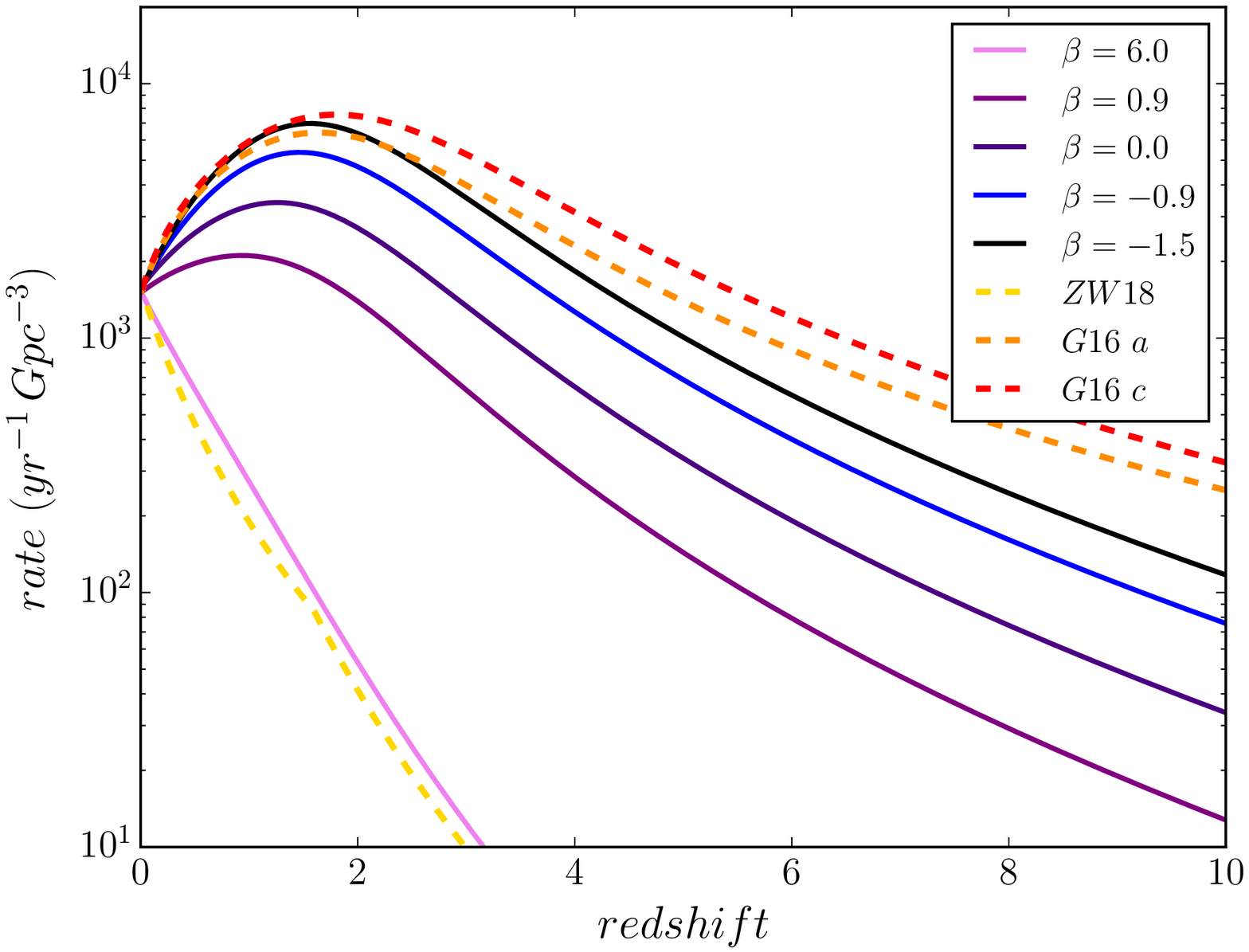}}
\subfloat[]{\includegraphics[width=0.5\textwidth]{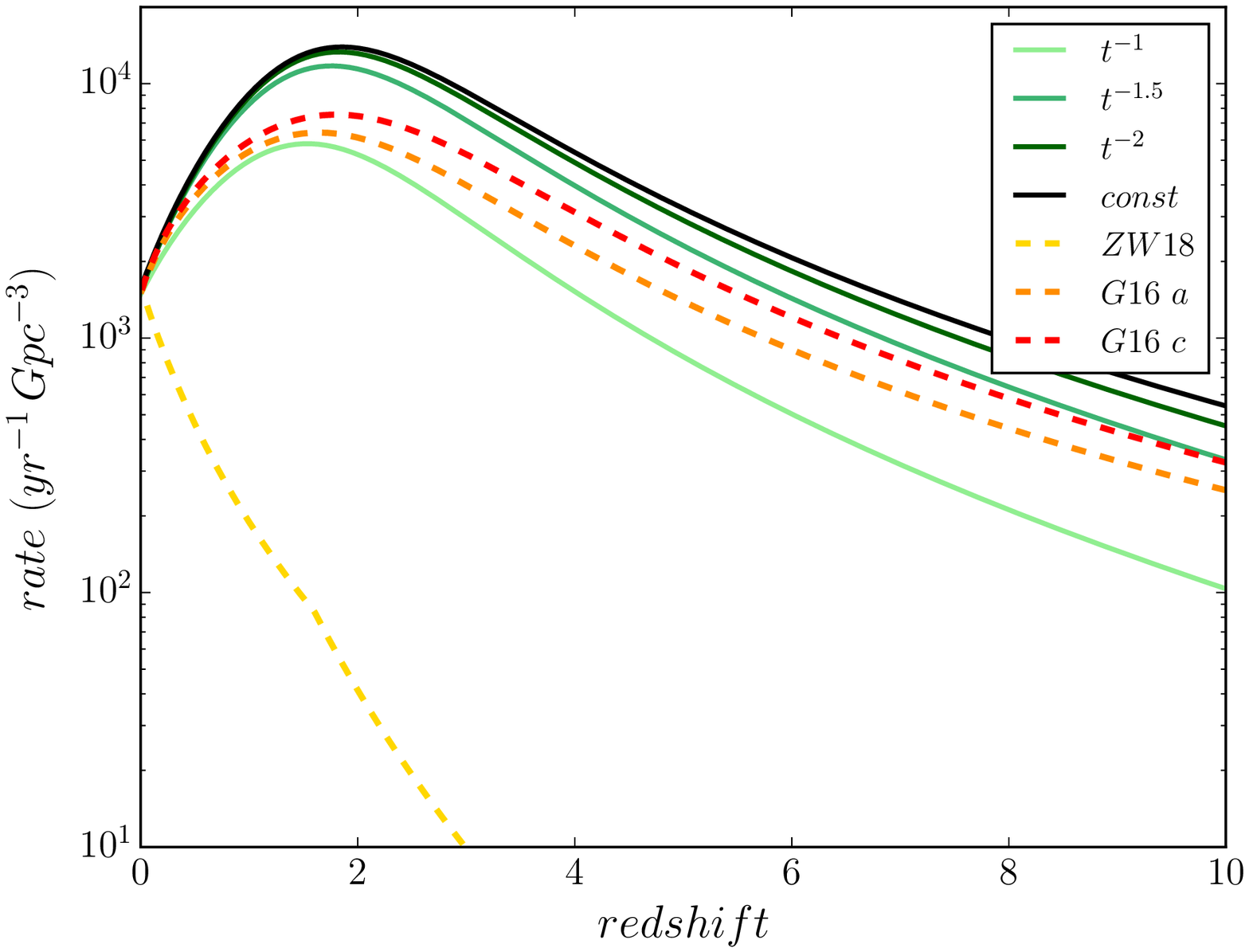}}
\caption{The redshift distribution of SGRB, as found by G16 (red and orange dashed lines) and ZW18 (yellow dashed line) against the predicted rate of MNS events (solid lines). Our DTD is shown in panel (a), with five different values for the $\beta$ parameter (pink=6.0; purple=0.9; violet=0.0; blue=-0.9; dark blue=-1.5). In panel (b) we show three different power laws ($\propto t^{-1}$ in light green; $\propto t^{-1.5}$ in green; $\propto t^{-2}$ in dark green) and a constant total delay of 10 Myr (in black).}
\label{cap5.DTDvSGRB}
\end{figure*}

The results can be seen in figure \ref{cap5.DTDvSGRB}. The rate of SGRB proposed by G16 is best represented by our DTD with $\beta=-1.5$, which means that a bottom heavy distribution (i.e. with many systems with small initial separations) is a good candidate. It is worth noting that such a value for $\beta$ gives rise to a distribution of time delays which scales as $\propto t^{-1.125}$. Lower values for $\beta$ imply higher and earlier maxima, so shorter time delays are preferred. However too short timescales, like those produced by the simple power law $\propto t^{-2}$ and the fixed 10 Myr total delay, do not produce a good agreement with the observations (see panel (b) of figure \ref{cap5.DTDvSGRB}: the maxima appear to exceed by a large factor the G16 distributions.)

The rate distribution proposed by ZW18, on the other hand, cannot be matched by any of the considered distributions. It requires a very top heavy distribution for the initial separations (with a $\beta=6.0$) and a low probability of forming close binaries ($\ams$ equal to $0.39 \times 10^{-2}$), but such a model is in strong contrast with the local data on chemical evolution, as it will be shown in the next section.

\section{Chemical evolution model for the Milky Way}

A chemical evolution model tracks the abundance of a given element $i$ in the gas of a galaxy at different times. The ingredients and the general theory of such a simulation can be found e.g. in \citet{matteucci12}. In particular, we have adopted a scheme similar to those described in \citet{matteucci14} that will be summarized below.

The evolution of the surface gas density of the element $i$, $G_i(r,t)$, is described by the following equation:
\begin{equation}
\label{sec3.gce}
\begin{split}
\frac{dG_i}{dt}(r,t)&= -\psi(r,t)Z_i(r,t)\,+ \\
&+\int_{m(t)}^{M_U} \psi(t-\tau(m))Q_{mi}(t-\tau(m))\phi(m)\,dm\,+ \\
&+Z_{i,0}A(r,t)
\end{split}
\end{equation}
The first term is the product of the star formation rate (SFR) $\psi(r,t)$ and the elemental fraction $Z_i(r,t)$ in the gas and represents the gas removed by the ongoing star formation. The second term represents the fraction of gas restored as element i by stars born at $t-\tau(m)$ and dying at the time $t$. $\tau(m)$ is the stellar lifetime for a star of mass $m$. In particular, $\phi(m)$ is the initial mass function (IMF) and $Q_{mi}(t-\tau(m))$ is the production matrix for a star of mass $m$ as defined by \citet{talbot73}. $M_U$ is the upper limit for the mass of a star, that we have chosen to be 100 $\ms$, while $m(t)$ is the minimum mass for a star dying at the time $t$. The mass dying at the present time is 0.8 $\ms$. The third term represents the infall of gas from outside the Galaxy. This infall follows a given rate in time at each galactocentric radius $A(r,t)$ and has a primordial (i.e. given by the Big Bang nucleosynthesis) composition, which for element $i$ is $Z_{i,0}$.

Our model follows the evolution of 31 chemical species, from H to Eu. The Milky Way is divided in concentric rings 2 kpc wide and eq. \eqref{sec3.gce} is integrated on a variable time step, whose minimum is 2 Myr. The output quantities represents the averages (for chemical abundances) or the totals (for rates) in the simulated Galaxy. Rates are calculated for MNS and SNeIa alike as in eq. \eqref{cap3.snrate}. We adopted the SFR by \citet{kennicutt98}, the IMF of \citet{kroupa93} and a double-exponential infall law \citep[the so-called \textit{two infall model} of][]{chiappini97} aimed at reproducing the present surface mass densities in the halo and the disk as observed by \citet{kuijken91}, with timescales from \citet{romano00}. No outflow is included and the instantaneous mixing approximation is retained \citep[because their delaying effect on the chemical evolution seems negligible, see e.g.][]{spitoni09}. The yields are taken from \citet{karakas10} for low and intermediate mass stars, from \citet{doherty14a,doherty14b} for super-AGB stars, from \citet{nomoto13} for SNeII, from \citet{iwamoto99} for SNeIa and from \citet{jose98} for nova systems. The stellar lifetimes are those of \citet{schaller92}.

\section{Comparison with the Galactic abundances of europium and iron}

In our calculation we need to specify the following parameters:
\begin{enumerate}[i.]
\item the DTD of the MNS;
\item the DTD of the SNeIa;
\item the production of Eu by CC-SNe;
\item the fraction of neutron stars which produce a MNS, introduced in section 2 ($\ams$);
\item the Eu produced per merging event.
\end{enumerate}
We have varied (i) and (ii) according to the options proposed in section 2. Concerning (iii), we have run two sets of scenarios, one in which CC-SNe are allowed to produce Eu and the other in which they are not. Finally, (iv) and (v) have been chosen according to the following observational constraints:
\begin{itemize}
\item $\ams$ has been fixed to the value found in section 3 to reproduce the cosmic LIGO/Virgo MNS rate \citep{abbott17a};
\item the yield of Eu has been fixed to reproduce the solar absolute abundance of Eu observed by \citet{lodders09}, in particular comparing it with the simulated abundance of Eu present in the Galaxy 9 Gyr after the Big Bang (which is about the age of formation of the Solar System). Given that different values of $\ams$ produce different predicted Eu abundances in the Sun, we have chosen a value that reproduces the observations in the case of intermediate timescales (models 4A and 4B in table \ref{cap5.models1}) and we have kept it constant. This choice makes easy to see the effects of different timescales on the total amount of Eu in the solar neighborhood.
\end{itemize}

\subsection{Eu production sites}

As discussed in the Introduction, there are two possible sites for Eu production: MNS and CC-SNe. In this paper, we have tested two different scenarios: one in which Eu is entirely produced by MNS and another in which Eu is produced both by MNS and CC-SNe.

\begin{figure}
\includegraphics[width=\columnwidth]{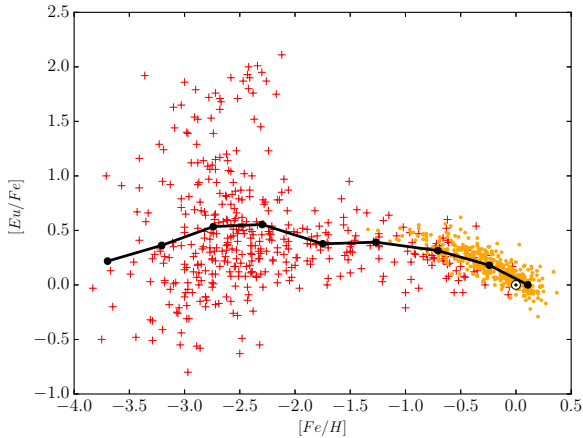}
\caption{Observational data used in paper: a compilation of 426 Milky Way halo stars (red crosses) taken from JINABase and 374 Milky Way thin disk stars (yellow points) from \citet{battistini16}. Black dots represent the average value, binned in 0.5 dex wide bins. It is possible to see the plateau at low metallicities ($\text{[Fe/H]}<-1.0$) and the decreasing trend at later times.}
\label{cap5.vuoto}
\end{figure}

MNS, as a production site of Eu, have been explored thoroughly by several authors \citep[e.g.][]{korob12} that proposed a yield in the range $10^{-7}-10^{-5}$ $\ms$ of Eu per event. More recently, the observation of the kilonova AT2017gfo has allowed us to estimate the yield of Eu in the range $(3-15)\times 10^{-6}$ $\ms$ of Eu per event \citep{evans17,tanvir17,troja17}. We have chosen to represent MNS as systems of two 1.4 $\ms$ neutron stars with progenitors in the 9-50 $\ms$ mass range.

We have chosen an empirical yield of Eu equal to $4.0 \times 10^{-6}$ $\ms$ when the MNS are the sole producers, and $1.5 \times 10^{-6}$ $\ms$ when Eu is co-produced by CC-SNe. This second value is slightly lower than the estimate from the kilonova AT2017gfo but well inside the theoretical range.

CC-SNe, on the other hand, have a less clear role in the r-process elements production. In this paper, we have used a slightly modified version of the provisions found in \citet{argast04}, model SN2050, and used also in the paper \citet{matteucci14}:
\begin{itemize}
\item a constant yield of $3.8 \times 10^{-8}$ $\ms$ ~of Eu for 20-23 $\ms$ ~mass range stars;
\item a decreasing yield from $3.8 \times 10^{-8}$ $\ms$ ~of Eu for a 23 $\ms$ ~star to $1.7 \times 10^{-9}$ $\ms$ ~of Eu for a 50 $\ms$ ~star.
\end{itemize}
We do not consider any dependence on metallicity of these yields, nor do we take into account the distribution in rotational velocities of high-mass stars and its effect on the yield of Eu.

If the timescale of Eu production were short, we would expect to see a plateau at low metallicities ($\text{[Fe/H]}<-1.0$) followed by a decline as the Fe abundance grows. This trend is found in the observational data (see figure \ref{cap5.vuoto}), and can be explained supposing that the main producers of Fe, namely the SNeIa, have lower mass progenitors than CC-SNe and longer gravitational delay times than MNS. Therefore, the bulk of Fe is released to the ISM on a longer timescale with respect to Eu. Initially, Eu and Fe are produced at a similar rate by only massive stars and this generates the plateau. Then, SNeIa start to explode, producing Fe but not Eu, progressively decreasing the [Eu/Fe]. This behaviour is similar to that of $\alpha$-elements (like O, Mg, Ca and Si) that are likewise produced in high mass short-lived stars, and is known as time delay model \citep[see,e.g.][]{matteucci12}. 

\subsection{Conversion of the cosmic rate to a Galactic rate}

As a consistency check, we have developed a simple conversion procedure to infer the Galactic MNS rate from from the cosmic LIGO/Virgo rate of \citet{abbott17a}. We have considered the total luminosity of a unitary volume of the Universe as derived by the integration of the Press-Schechter function, equal to $2\times 10^{17}\times h$ $L_{\odot}$, as reported by \citet{mo10}. Here, $h$ is the dimensionless parametrization of the Hubble constant in units of 100 $\text{km}\,\text{s}^{-1}\,\text{Mpc}^{-1}$. 

Considering a total baryonic mass for our Galaxy equal to $(5-10) \times 10^{10}$ \citep[as found by e.g.][]{sofue13,mcmillan16} and a mass-to-luminosity ratio for a spiral galaxy equal to 5 \citep[][]{mo10}, we have estimated the total luminosity of the Milky Way. The result is that there are $\sim (1-2) \times 10^{7} \times h$ Milky Way-equivalent (MWe) galaxies per $\text{Gpc}^3$. Choosing a value for $h$ equal to 0.7 as most cosmological observations suggest \citep{planck16} produces a central value for the obtained interval of $\sim 1 \times 10^{7}$ $\text{MWe}\,\,\text{Gpc}^{-3}$. Therefore, we expect that the galactic rate of MNS is $\simeq 10^{-7}$ times lower than the \citet{abbott17a} value, or $1.54_{-1.22}^{+3.2} \times 10^{-4}$$\text{yr}^{-1}$. This figure compares well to previous Galactic estimations based on binary pulsars such as $0.83_{-0.7}^{+2.1} \times 10^{-4}$ $\text{yr}^{-1}$ \citep{kalogera04}. In our chemical evolution model the current rate of MNS is not a free parameter, but results from the convolution of the star formation history of the Milky Way with the DTD, modulo the fraction $\alpha_{NMS}$ derived from the cosmic constraint. We will compare the model results to these figures.

\subsection{Simulations and results}

We ran several models, as reported in table \ref{cap5.models1}. In the first column we put the name of the model, in the second column is indicated if the CC-SNe have been considered as production site for Eu or not and in the third and fourth columns are indicated the DTDs used, respectively, for SNeIa and MNS. In the fifth and sixth columns are reported the values for the occurrence probabilities of SNeIa and MNS. The occurrence probabilities $\alpha_{Ia}$ have been chose so as to obtain the estimated current rate of Type Ia supernovae, $1.8 \times 10^{-3}\, S\!N\, yr^{-1}$ in the Milky Way \citep{li11}, while $\alpha_{MNS}$ has been fixed by the fit of the local volumetric SGRB rate (see Sect. 3). In the seventh column is reported the adopted Eu yield per MNS event, in the eighth column the predicted current rate of MNS in the Milky Way and in the ninth and tenth columns are reported the predicted absolute solar abundances of Fe and Eu.

\begin{figure}
\includegraphics[width=\columnwidth]{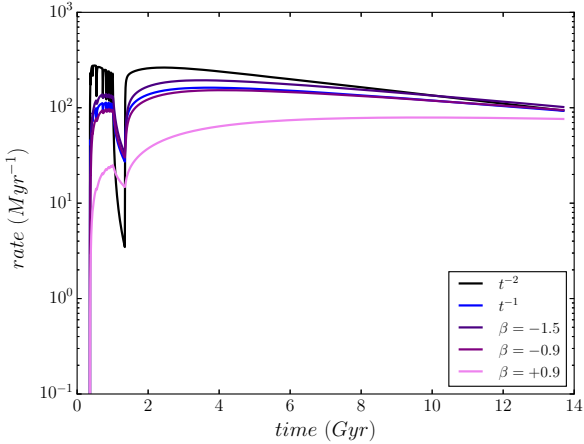}
\caption{The Galactic rate of MNS for five of the eight DTDs tested here: 1A/B in pink, 3A/B in violet, 4A/B in indigo, 5A/B in blue and 7A/B in black.}
\label{sec4.rates}
\end{figure}

As it is possible to see in table \ref{cap5.models1}, the current galactic MNS rate turns out smaller than the estimate in Sect. 5.2 obtained extrapolating the \citet{abbott17a} value, but it appears to be in very good agreement with \citet{kalogera04} estimate. Given that Galactic and cosmic star formation histories are different we do expect some differences in the current MNS rate, at fixed DTD and $\alpha_{MNS}$. The effect of the Galactic SFR on the MNS rate can be clearly seen in figure \ref{sec4.rates}, where the computed MNS rates are shown: at around 1 Gyr from the formation of the Galaxy, when the first infall episode terminates, the merging rate rapidly decreases. The magnitude of such a decrease depends on the timescale of the adopted DTD: longer ones act as smoothing masks. The rapid oscillations at very early times are caused by the threshold in the SFR adopted in the chemical model, as described in Sect. 4.

As a first step, it has been verified the ability of different DTDs to reproduce the solar abundances of Eu and Fe.

For the Sun, Lodders et al. (2009) have determined:
\[
\log(Fe/H)_{\odot}=-2.792 \quad \log(Eu/Fe)_{\odot}=-6.496
\]
\[
X_{E\!u}=3.50 \times 10^{-10} \quad X_{F\!e}=1.34 \times 10^{-3}.
\]
These values should be compared with the predicted ones in table \ref{cap5.models1}.

We have kept the same yields of Fe for SNeII and SNeIa during all the tests and instead varied the DTD, using the two possible distributions listed in section 5.1. The values reported in Table \ref{cap5.models1} show that the solar Fe abundance in our model is in excellent agreement with the observational value. For the Eu contribution from MNS, the yields were actually tuned to reproduce the solar abundance with our model. We can see that the required production is quite compatible with theoretical estimates, as well as the empirical value from of the kilonova, as mentioned in Sect. 5.1.

\begin{figure*}
\centering
\subfloat[]{\includegraphics[width=0.5\textwidth]{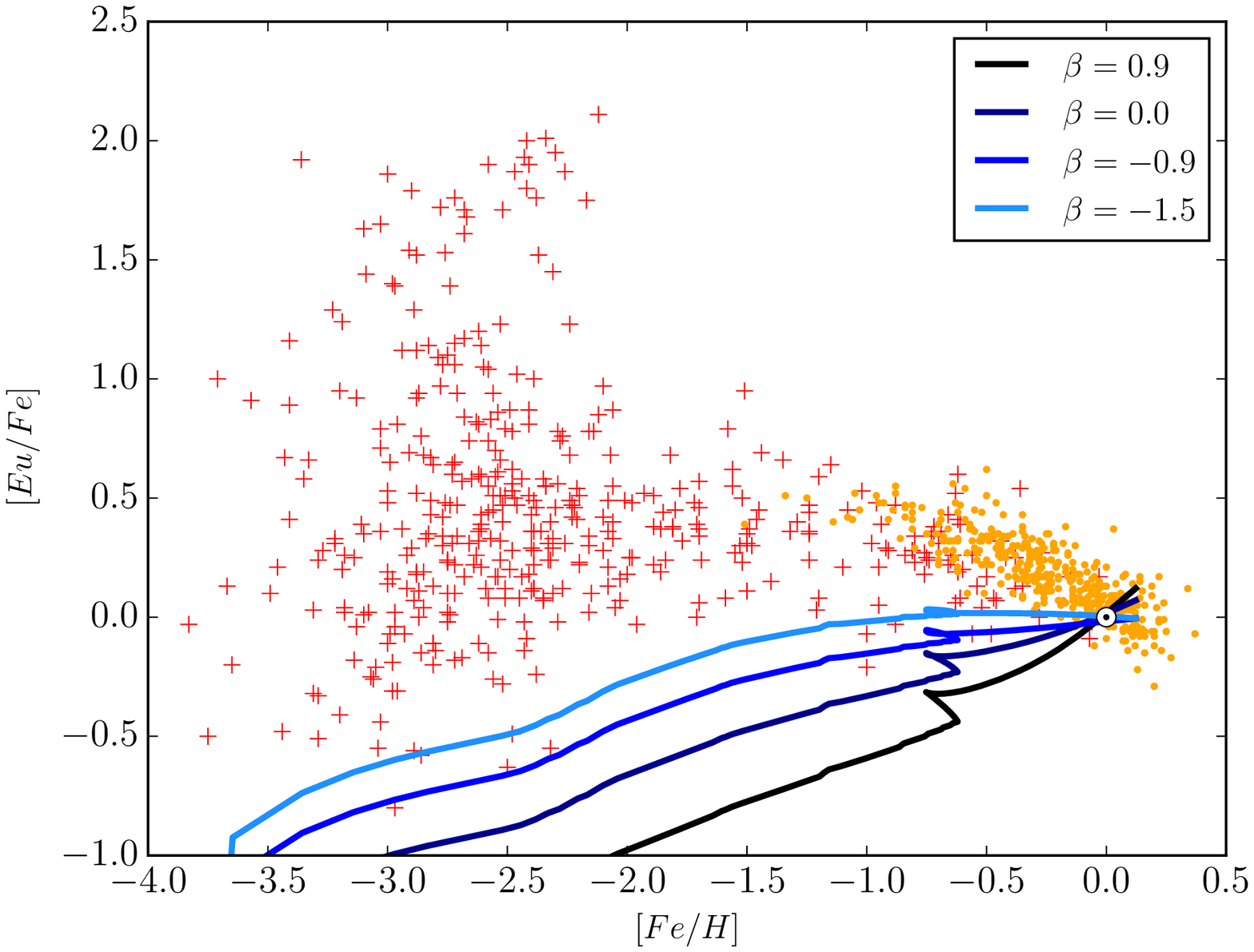}}
\subfloat[]{\includegraphics[width=0.5\textwidth]{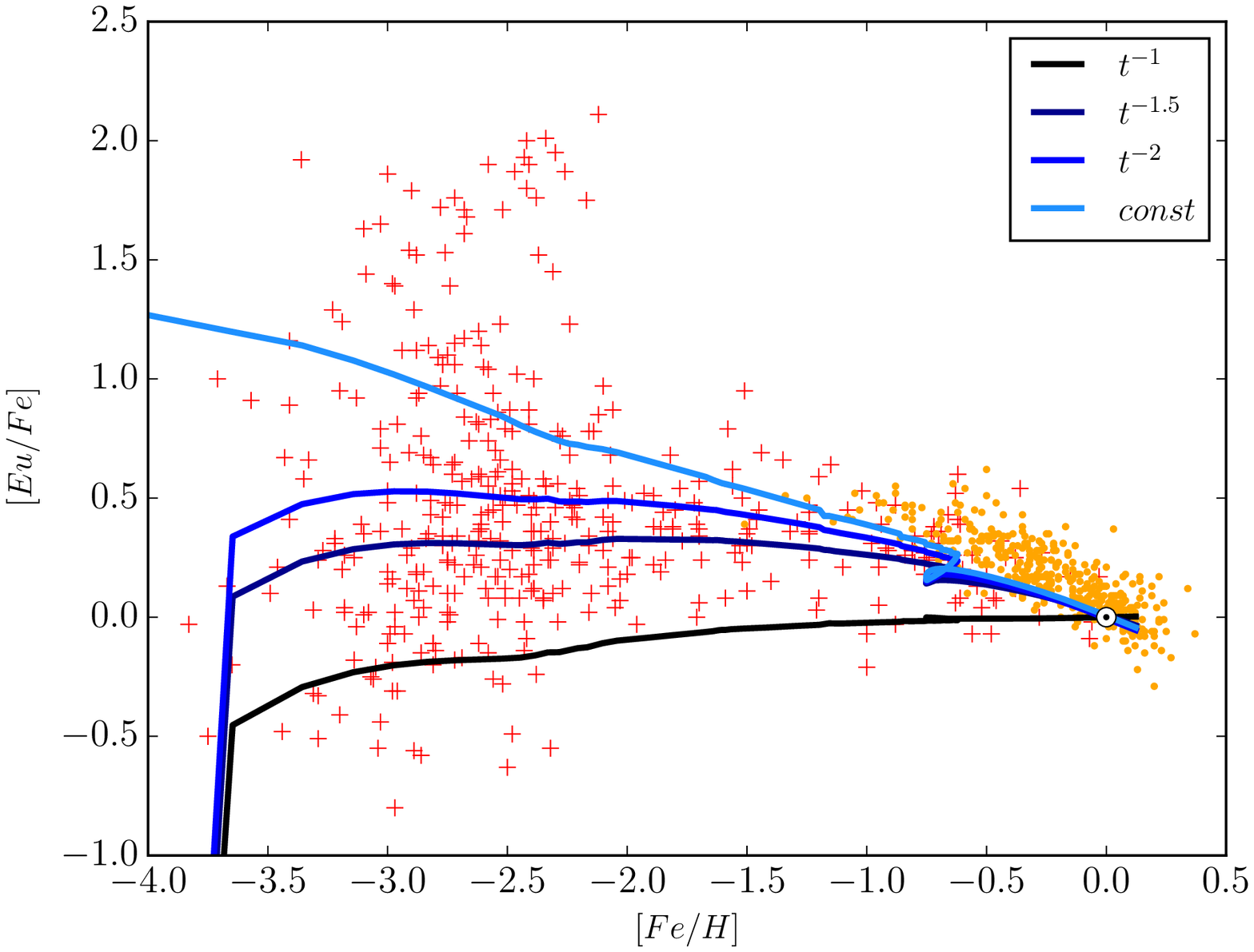}}\\
\subfloat[]{\includegraphics[width=0.5\textwidth]{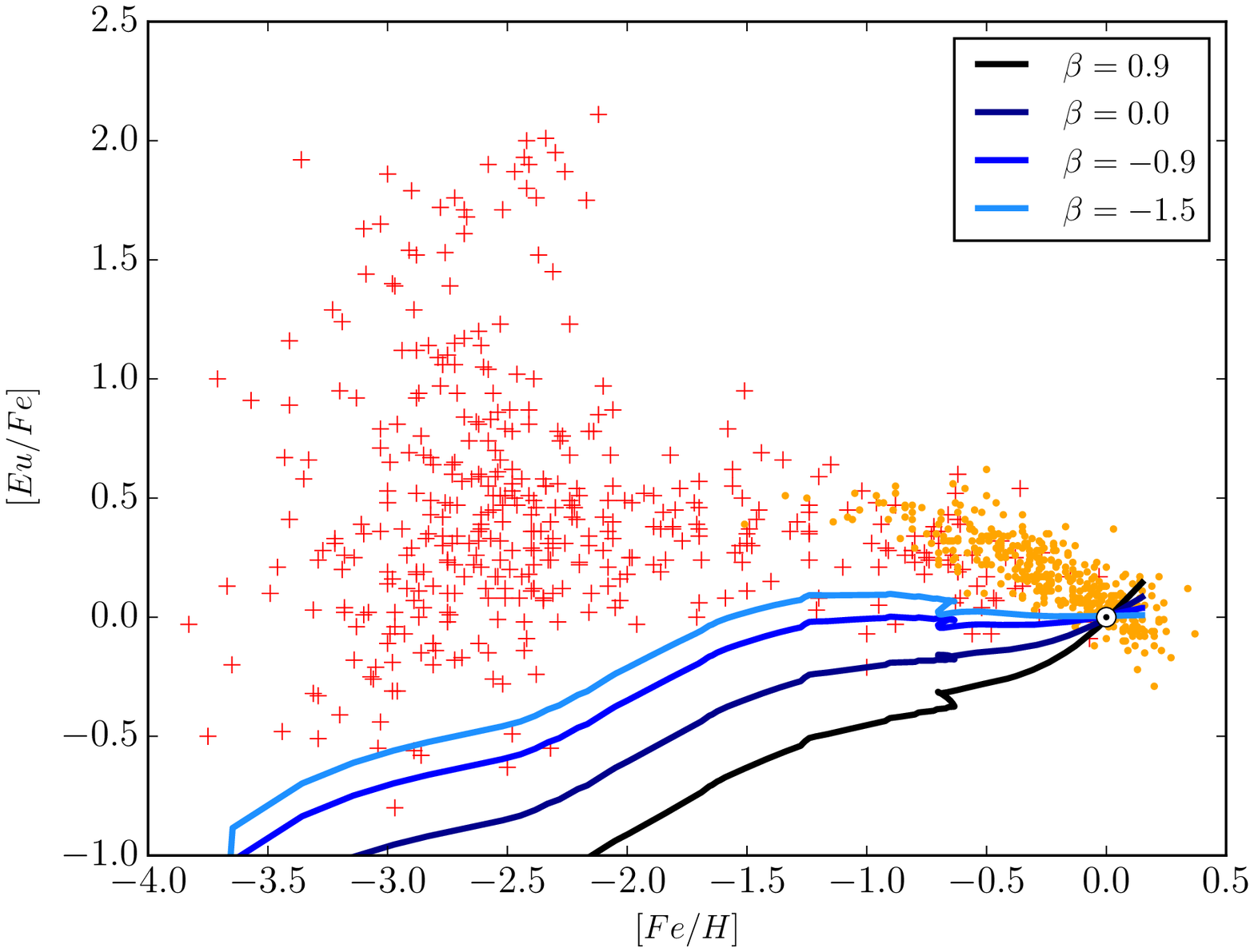}}
\subfloat[]{\includegraphics[width=0.5\textwidth]{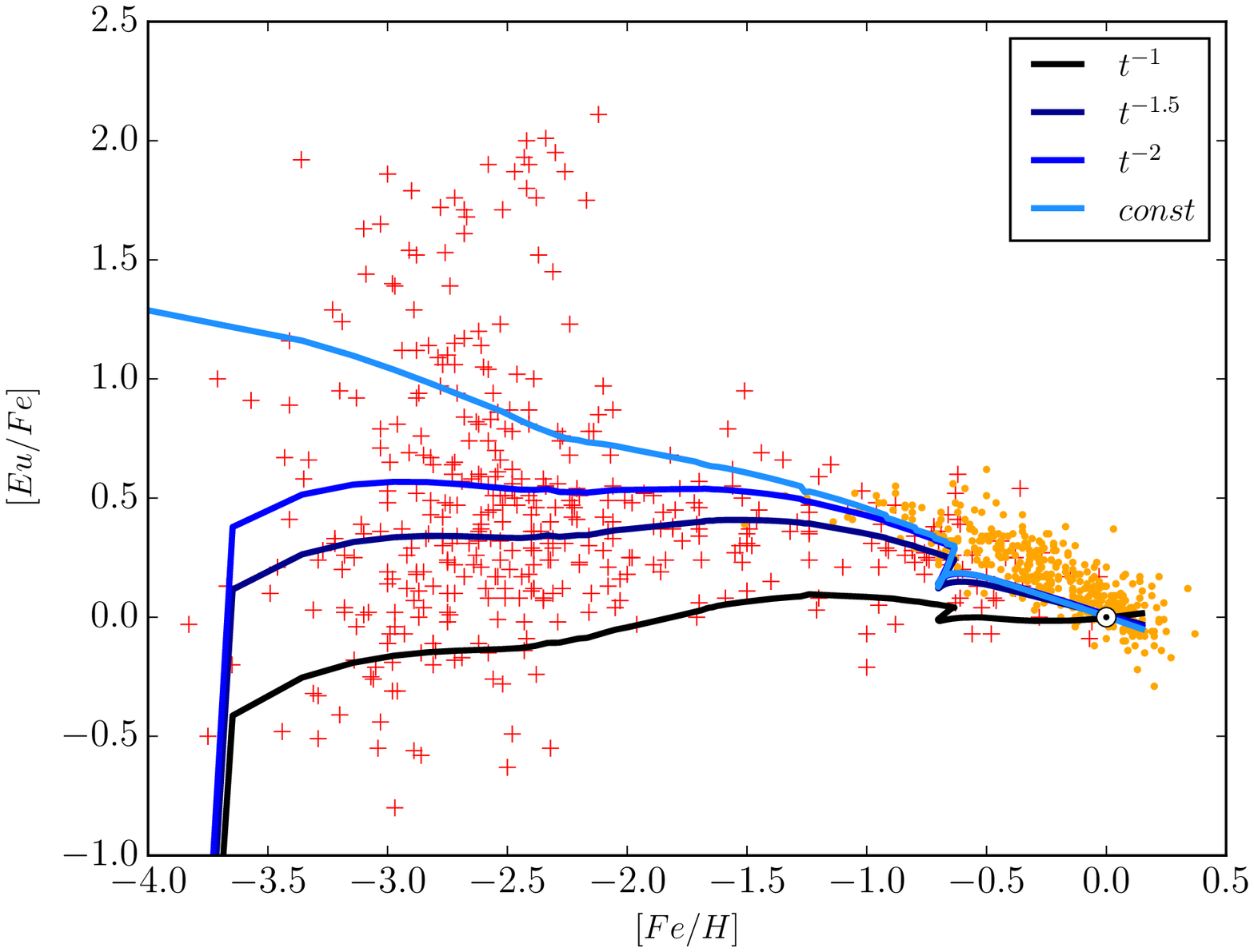}}
\caption{Here we show the models with no Eu production from CC-SNe. Panel (a) shows the models from 1A to 4A, panel (b) the models from 5A to 8A, panel (c) the models from 1B to 4B and panel (d) the models from 5B to 8B. The color-code is the same everywhere: lighter shades of blue stands for higher-index models.}
\label{sec5.noccsne}
\end{figure*}

Now we can turn to study the [Eu/Fe] vs [Fe/H] relation. We remind that the [X/Y] notation refers to the logarithm of the ratio between the two elements X and Y, normalized with respect to the same logarithmic ratio in the Sun. In figure \ref{sec5.noccsne} we show the effects of different DTDs on the [Eu/Fe] vs [Fe/H] relation when CC-SNe are not allowed to produce Eu. In particular, we are interested in reproducing the decreasing trend of [Eu/Fe] vs [Fe/H] in the disk stars (i.e. those at $\text{[Fe/H]}>-1.0$). We will not study the spread in the datapoints in the halo stars as done by other authors \citep[e.g.][]{cescutti15,wehmeyer15}, since our model is homogeneous. This spread is likely due to local polluting events.

We can categorize the models in three groups: those that totally fail to produce the decreasing trend in the disk stars, those that produce an insufficient decreasing trend and those that produce an acceptable decreasing trend. The models in the first category usually fails also to reproduce the low metallicity plateau. As it is possible to see in panels (a) and (c), the DTD derived in this paper falls in the first category, meaning that coalescence timescales longer than 300 Myr are unable to reproduce the [Eu/Fe] vs [Fe/H] relation along the entire range of metallicities. 

On the other hand, three out of four models shown in panels (b) and (d) reproduce the decreasing trend at high metallicity. The models that produce the best correspondence are the 7A/B and the 8A/B. In particular, the 7A/B also reproduce the plateau at $\text{[Fe/H]}<-1.0$ and can be selected as the best model. Therefore, we can say that a DTD $\propto t^{-2}$ is adequate to reproduce the [Eu/Fe] abundance pattern in the Milky Way stars, if MNS are the only contributors to Eu.

\begin{figure*}
\centering
\subfloat[]{\includegraphics[width=0.5\textwidth]{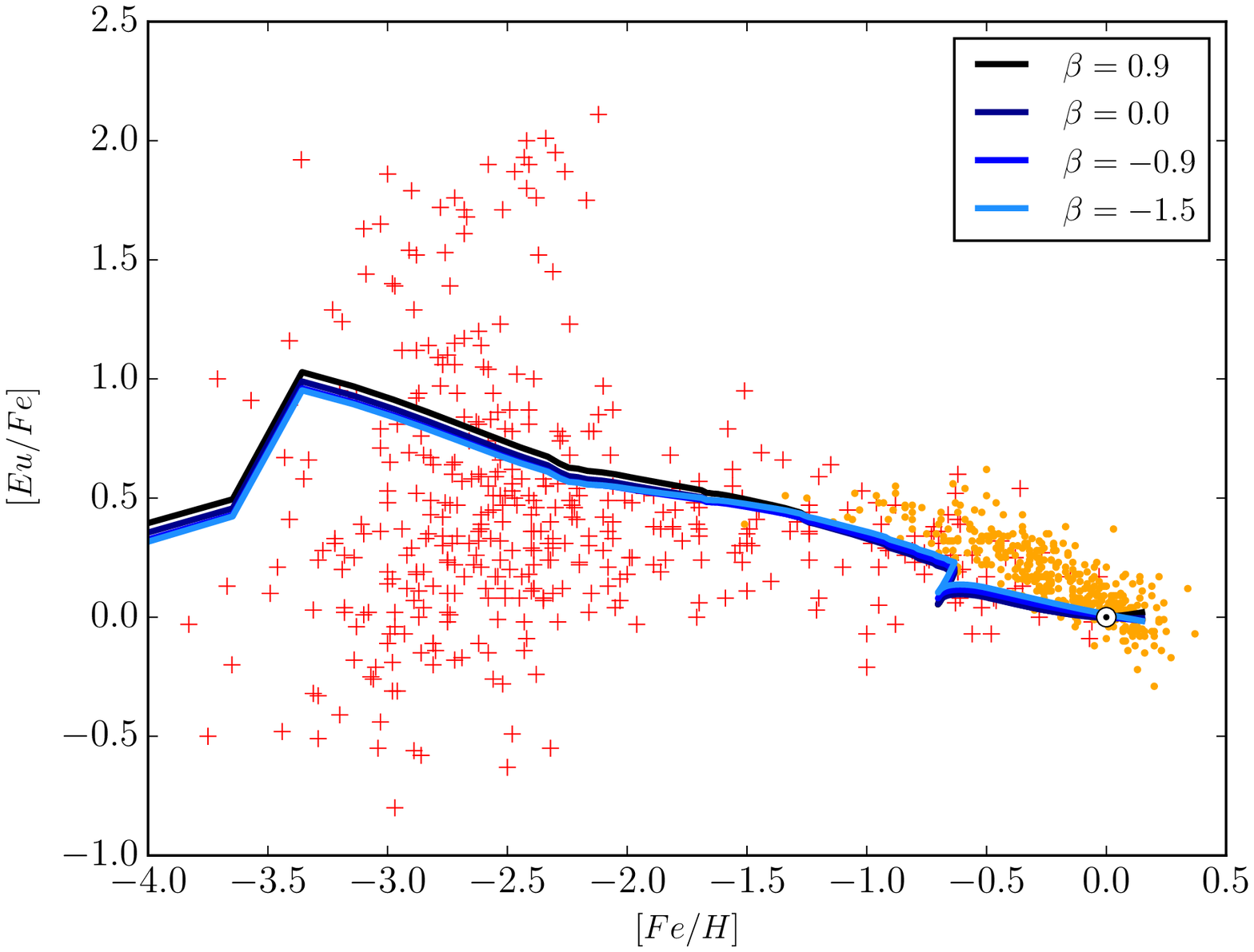}}
\subfloat[]{\includegraphics[width=0.5\textwidth]{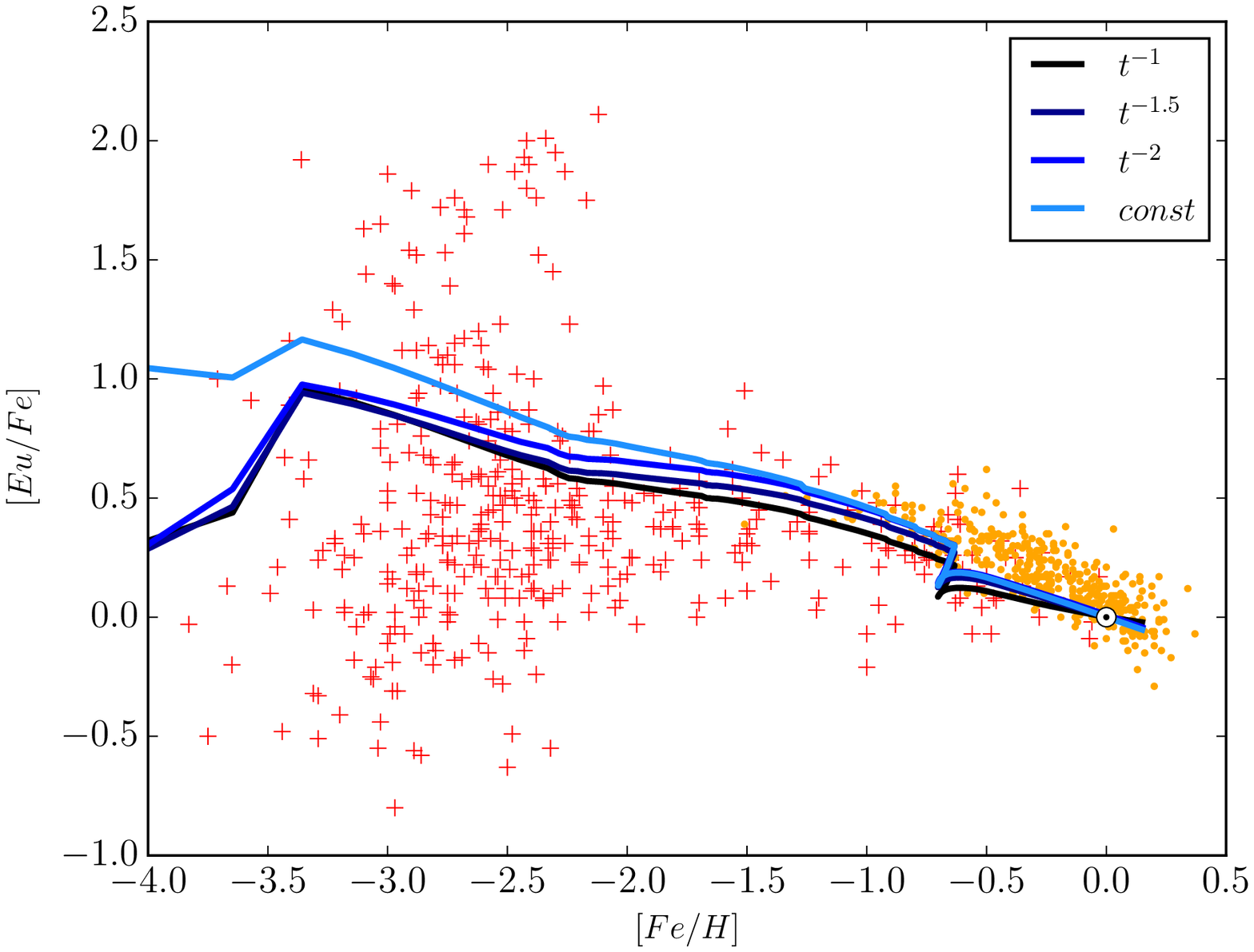}}
\caption{Here we show the models with Eu production from both CC-SNe and MNS. Panel (a) shows the models from 1BS to 4BS and panel (b) shows the models from 5BS to 8BS. The color-code is the same everywhere: lighter shades of blue stands for higher-index models.}
\label{sec5.ccsne}
\end{figure*}

In figure \ref{sec5.ccsne} we show the models where CC-SNe co-produce Eu alongside MNS. The impact of the CC-SNe contribution is very large since with the prescriptions used here no less than 60\% of the solar Eu should come from CC-SNe. Here, all the DTDs produce a decreasing trend in the [Eu/Fe] vs [Fe/H] relation. In particular, DTDs with longer timescales (like the one derived in this paper) become indistinguishable from one another (see panel a). For shorter timescales (panel b) there are slight differences between the models. One more time, shorter DTDs (7BS, 8BS) are favoured over longer ones (5BS). The low metallicity plateau, although deformed by the knee at $\text{[Fe/H]} \sim -3.3$, is visible nonetheless. Such a knee is caused by the death of massive stars in the lower region ([20-23] $\ms$) of the allowed mass range, where the yields of Eu are higher. We do not show the graphs from the 1AS to the 8AS model because they are substantially equal to those shown in figure \ref{sec5.ccsne}.

\begin{figure*}
\centering
\subfloat[]{\includegraphics[width=0.5\textwidth]{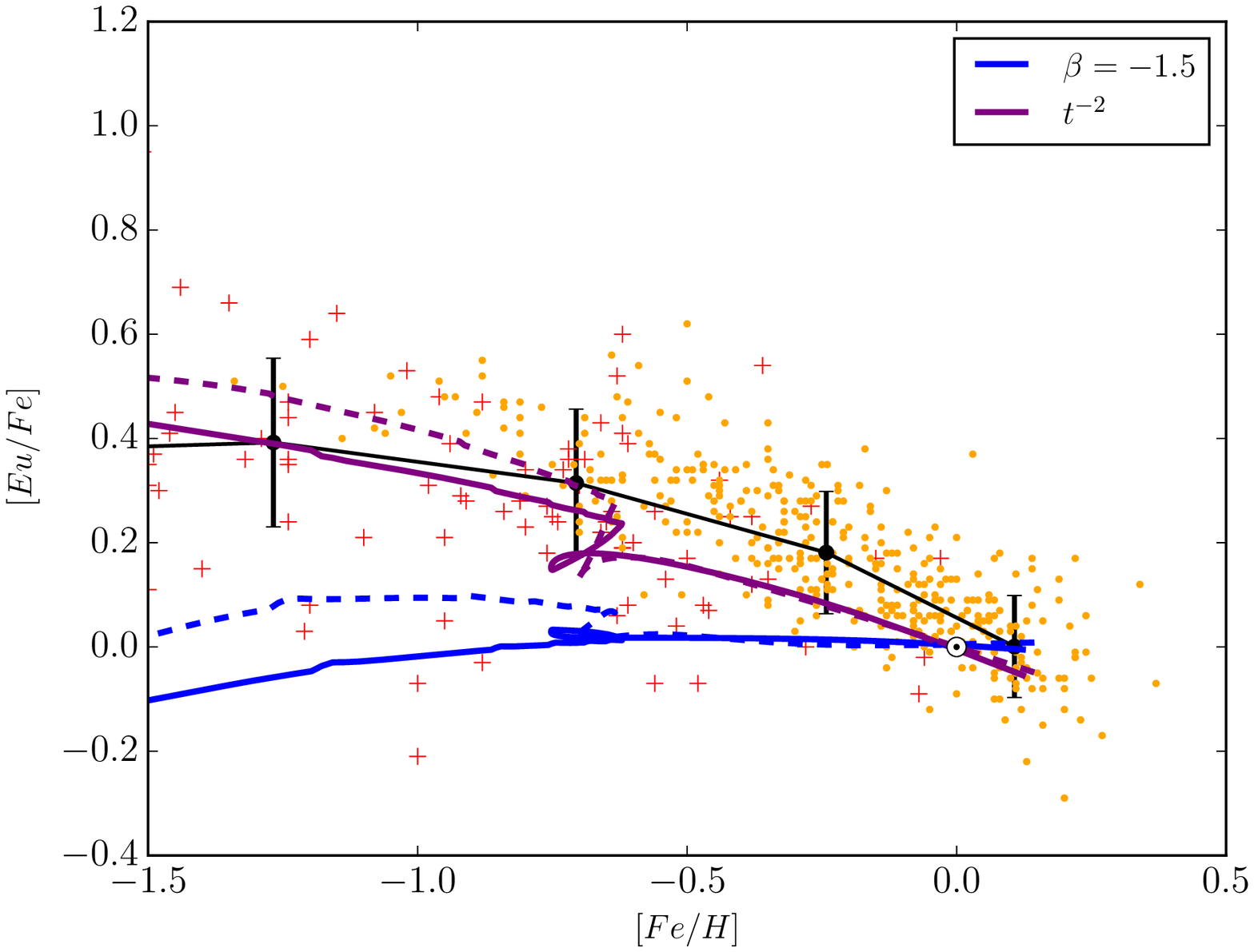}}
\subfloat[]{\includegraphics[width=0.5\textwidth]{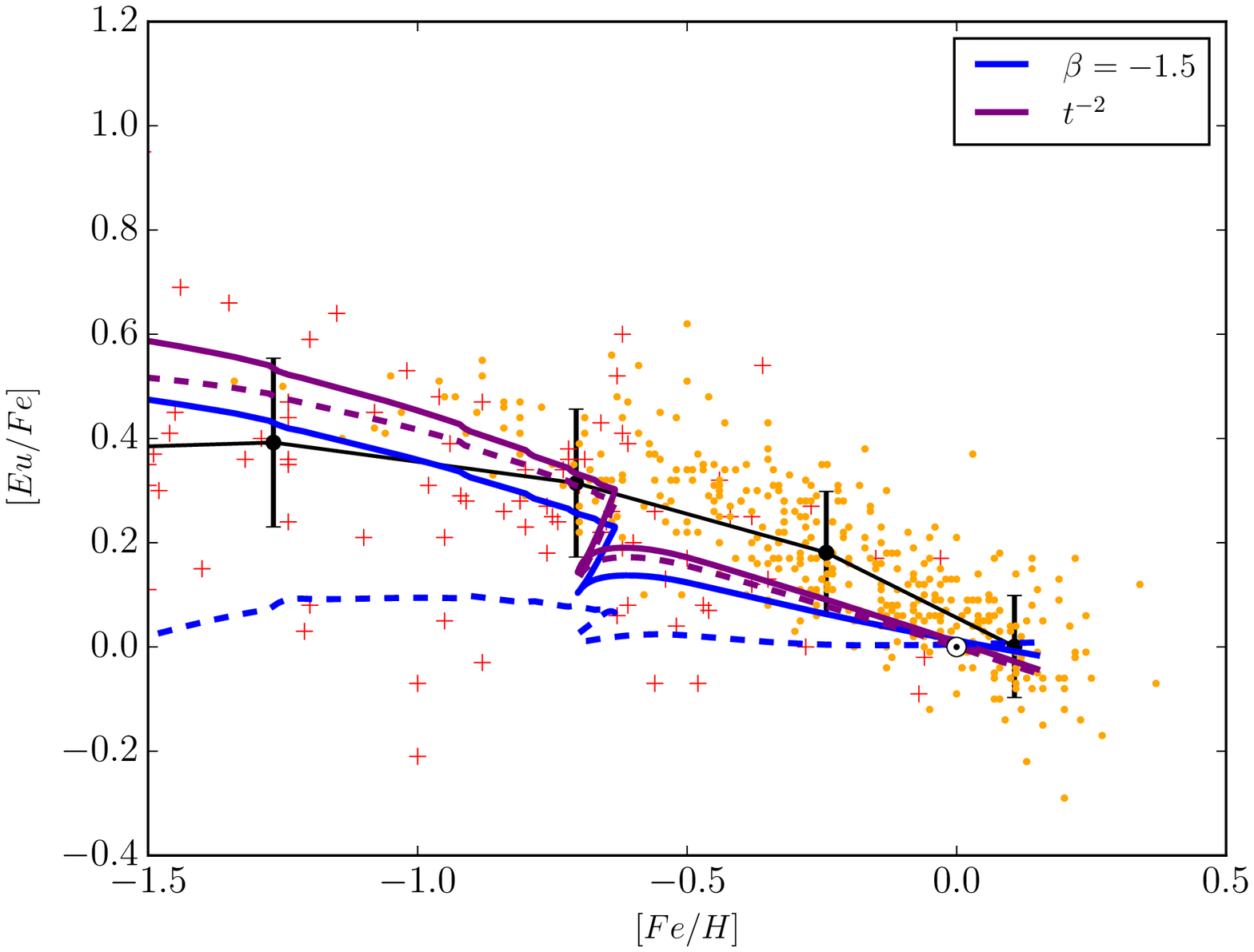}}
\caption{In panel (a) we highlight what happens when we change DTD for SNeIa. Blue lines refer to our DTD with $\beta=-1.5$ (models 4A/B), purple lines refer to the DTD $\propto t^{-2}$ (models 7A/B). Solid lines represent models with a DTD for SNeIa $\propto t^{-1}$ (models 4/7A), while dashed lines represent models with a G05 DTD for SNeIa (models 4/7B). In panel (b) we highlight what happens when we activate the production of Eu in CC-SNe. Solid lines represent models where Eu is co-produced by CC-SNe (models 4/7BS), while dashed lines represent models where MNS are the sole producers of Eu (models 4/7B). Black points represents the average [Eu/Fe] in a 0.5 dex wide bin and error bars are 1-$\sigma$ tall.}
\label{sec5.confront}
\end{figure*}

In figure \ref{sec5.confront} we show in detail the effect of changing DTD for SNeIa (panel a) and activating/deactivating Eu production in CC-SNe (panel b) in the abundance range [Fe/H] > -1.5. Using the G05 DTD for SNeIa (panel a, dashed lines), that produces longer timescales than $\propto t^{-1}$, slightly increase the steepness in $\text{[Fe/H]}<-0.5$, while has substantially no impact in the $\text{[Fe/H]}>-0.5$. However, it is not sufficient to make our DTD (with $\beta=-1.5$) acceptable. Instead, the DTD $\propto t^{-2}$ offers an optimal profile in both cases (models 7A and 7B). If we observe panel (b), we can see that the activation of Eu production in CC-SNe strongly improves the prediction when using our DTD (model 4BS), even if it remains worse than in the case of shorter timescales.

\begin{figure*}
\centering
\subfloat[]{\includegraphics[width=0.5\textwidth]{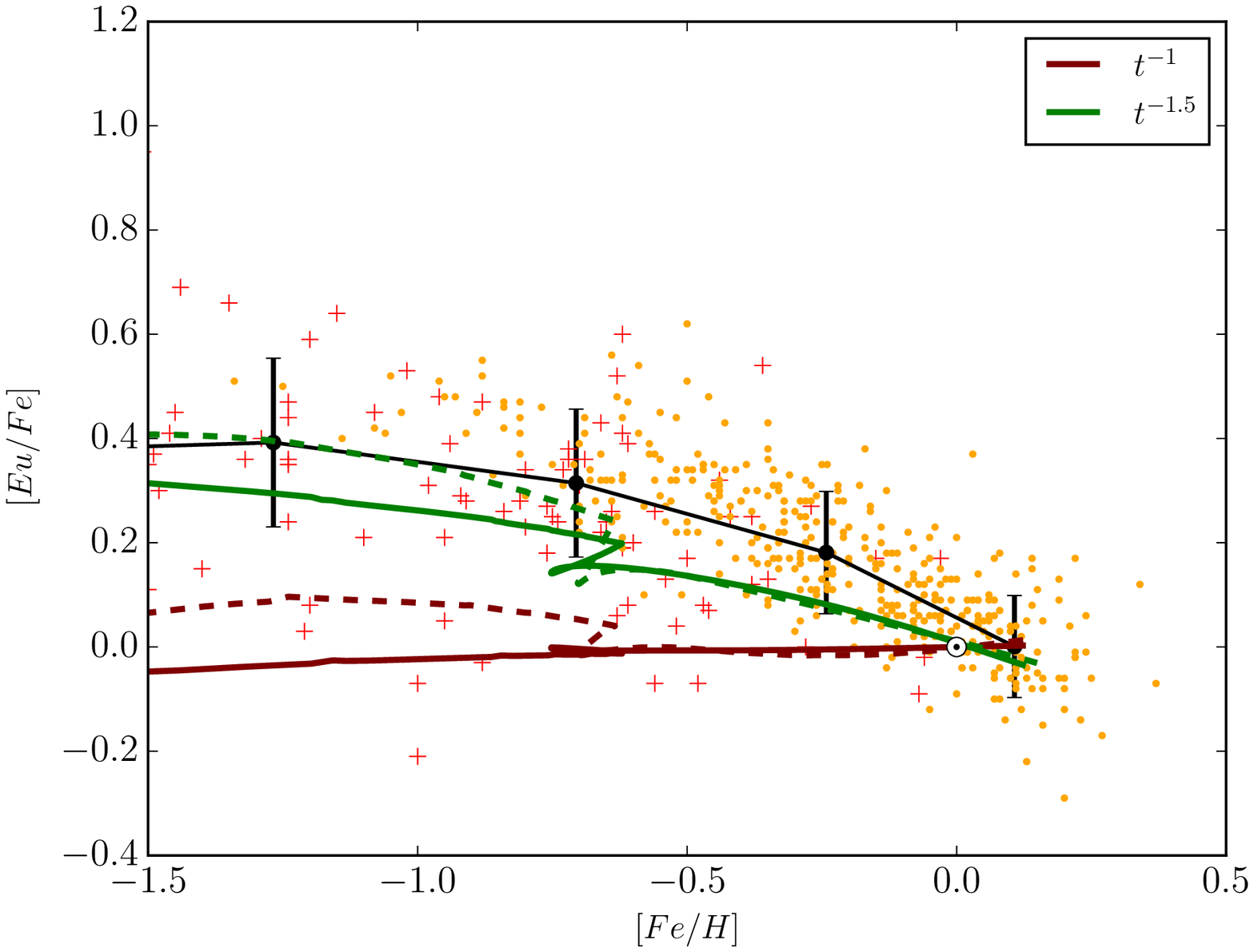}}
\subfloat[]{\includegraphics[width=0.5\textwidth]{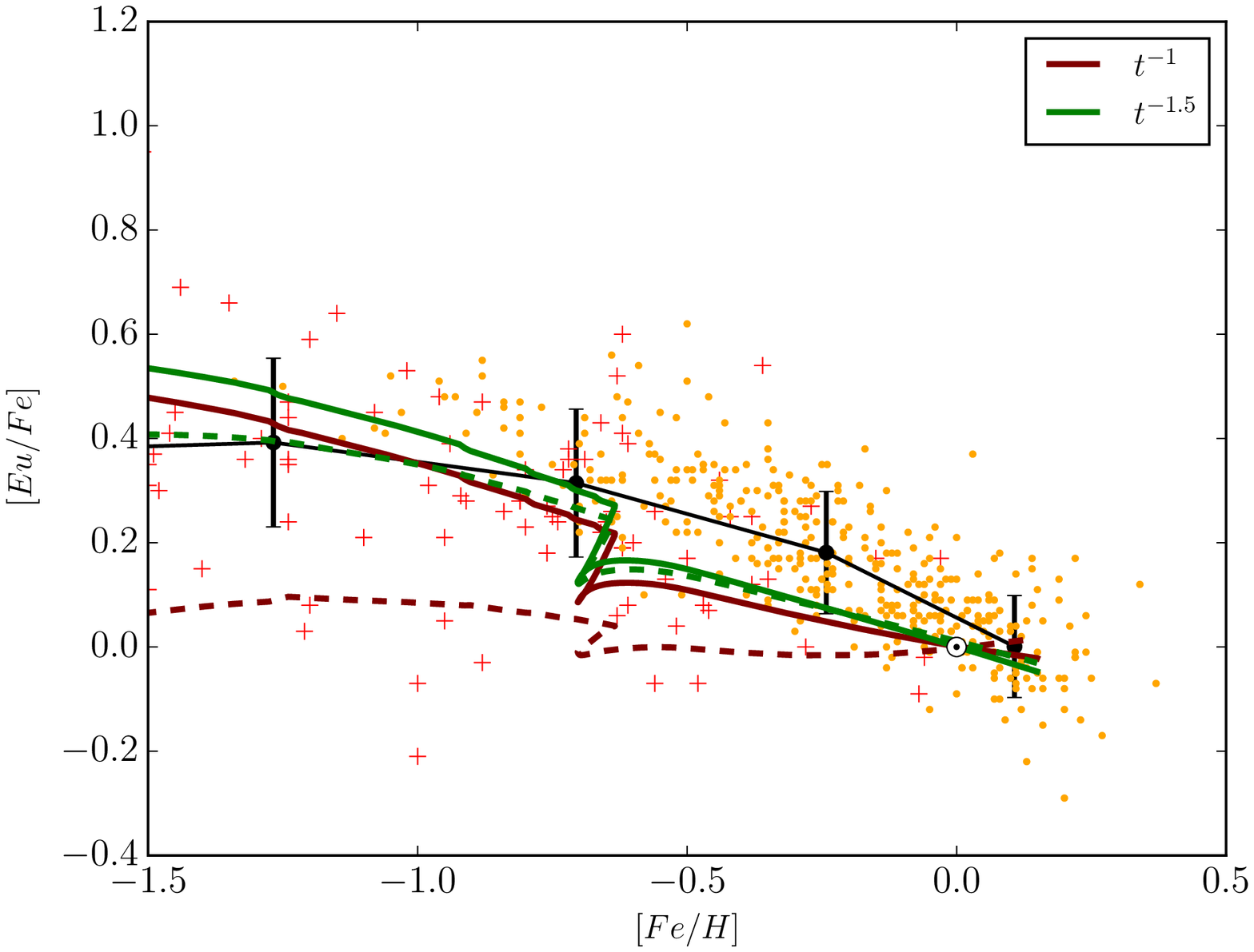}}
\caption{In panel (a) we highlight what happens when we change DTD for SNeIa. Brown lines refer to the DTD $\propto t^{-1}$ (models 5A/B/BS), green lines refer to the DTD $\propto t^{-1.5}$ (models 6A/B/BS). Solid lines represent models with a DTD for SNeIa $\propto t^{-1}$ (models 5/6A), while dashed lines represent models with a G05 DTD for SNeIa (models 5/6B). In panel (b) we highlight what happens when we activate the production of Eu in CC-SNe. Solid lines represent models where Eu is co-produced by CC-SNe (models 5/6BS), while dashed lines represent models where MNS are the sole producers of Eu (models 5/6B). Black points represents the average [Eu/Fe] in a 0.5 dex wide bin and error bars are 1-$\sigma$ tall.}
\label{sec5.confront2}
\end{figure*}

Finally, in figure \ref{sec5.confront2} we show the plots relative to two popular choices for MNS: the DTD $\propto t^{-1.5}$ and the DTD $\propto t^{-1}$. We recall from section 4 that, even if they are not the preferred models to reproduce the SGRB redshift distribution, they cannot be entirely ruled out. As we can see, we confirm what has been found by other studies \citep[e.g.][]{cote17b}: a DTD $\propto t^{-1}$ is unable to recover the decreasing trend in the thin disk abundance pattern (panel a). If we activate the Eu production in CC-SNe (panel b), we obtain a relation that, although decreasing, is not sufficiently steep. A more satisfactory prediction is given by the DTD $\propto t^{-1.5}$ (models 6A/B and 6AS/BS), both with and without CC-SNe producing Eu. However, the steepness of the curve is one more time smaller than the observed one.

\subsection{A variable $\ams$}

\begin{figure}
\includegraphics[width=\columnwidth]{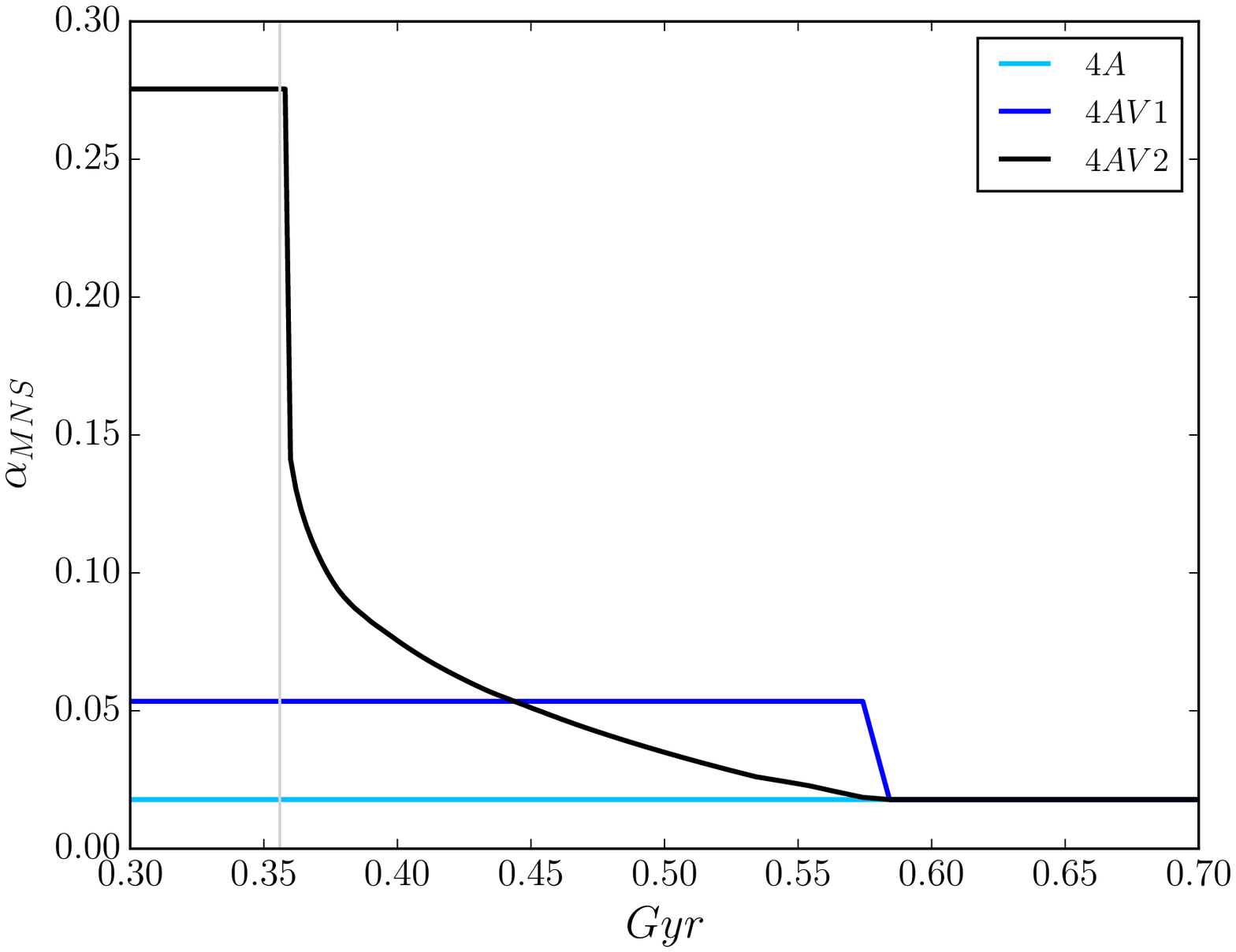}
\caption{The evolution of the $\ams$ parameter in three different scenarios. The gray vertical line represents the beginning of star formation in our simulation of the Milky Way.}
\label{sec5.variants}
\end{figure}

CC-SNe seem to provide a way to (partially) reconcile local and cosmic data. Another possible choice is to relax the assumption of constancy of the occurrence probability of close binary neutron stars $\ams$. This parameter depends on the physics of the stellar formation and the efficiency of the common envelope phase in the late stages of stellar life. Population synthesis models \citep[e.g][]{giacobbo18a} indicate that metallicity plays an important role. Therefore, we tested a variation  of the MNS occurrence frequency in the best fit model for the SGRB redshift distribution (the 4A, i.e. the one that uses the DTD derived in this paper with $\beta=-1.5$). In particular, we assumed that $\ams$ depends on [Fe/H], called $Z_{F\!e}$ from now on. All the other parameters are left the same, as reported in table \ref{cap5.models1}, and the value for $\ams$ reported in the table for the model 4A will be indicated as $\tilde{\alpha}_{M\!N\!S}$. We tested two dependencies of $\ams$ on $Z_{F\!e}$:
\begin{enumerate}[i.]
\item model 4AV1, where $\ams$ varies as:
\begin{equation}
\ams(Z)=
\begin{cases}
3\tilde{\alpha}_{M\!N\!S} & Z_{F\!e}\le-1.0\\
\tilde{\alpha}_{M\!N\!S} &  Z_{F\!e}>-1.0
\end{cases}
\end{equation}
\item model 4AV2, where $\ams$ varies as:
\begin{equation}
\ams(Z)=
\begin{cases}
\tilde{\alpha}_{M\!N\!S}(1+5\ln(-Z)) & Z_{F\!e}\le-1.0\\
\tilde{\alpha}_{M\!N\!S} &  Z_{F\!e}>-1.0
\end{cases}
\end{equation}
\end{enumerate}
The evolution of the parameter $\ams$ is reported in figure \ref{sec5.variants}. For the primordial gas, the absolute abundance of iron has been set to zero, and the initial value for $\ams$ of 27.5 \% has been chosen in order to best fit the observational data. Then, the sharp decrease of $\ams$ in the model 4AV2 (black line) is produced by the rapid increase in metallicity caused by the death of the very first stars. This procedure is somewhat arbitrary but it is useful to show which variation of $\ams$ can reconcile the SGRB cosmic rate with the [Eu/Fe] vs. [Fe/H].

\begin{figure}
\includegraphics[width=\columnwidth]{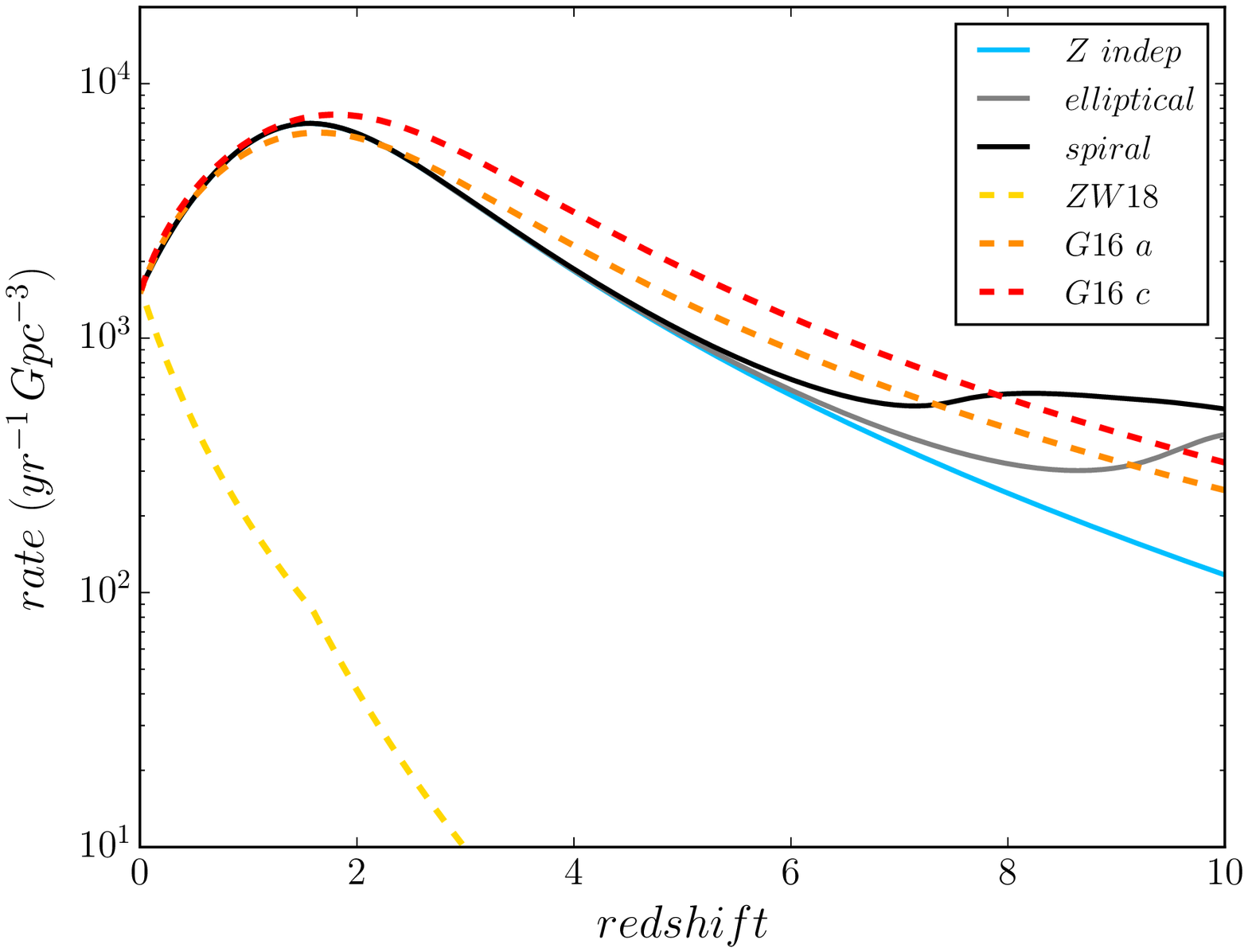}
\caption{The predicted evolution of SGRB rate against the observed ones. The models shown are the 4A (in light blue) and the 4AV2 (black and gray lines). In particular, the gray line refers to the sub-model with a [Fe/H] evolution as predicted for an elliptical galaxy, whereas the black line refers to the sub-model with a [Fe/H] evolution as predicted for the Milky Way.}
\label{sec5.SGRBvar}
\end{figure}

\begin{figure*}
\centering
\subfloat[]{\includegraphics[width=0.5\textwidth]{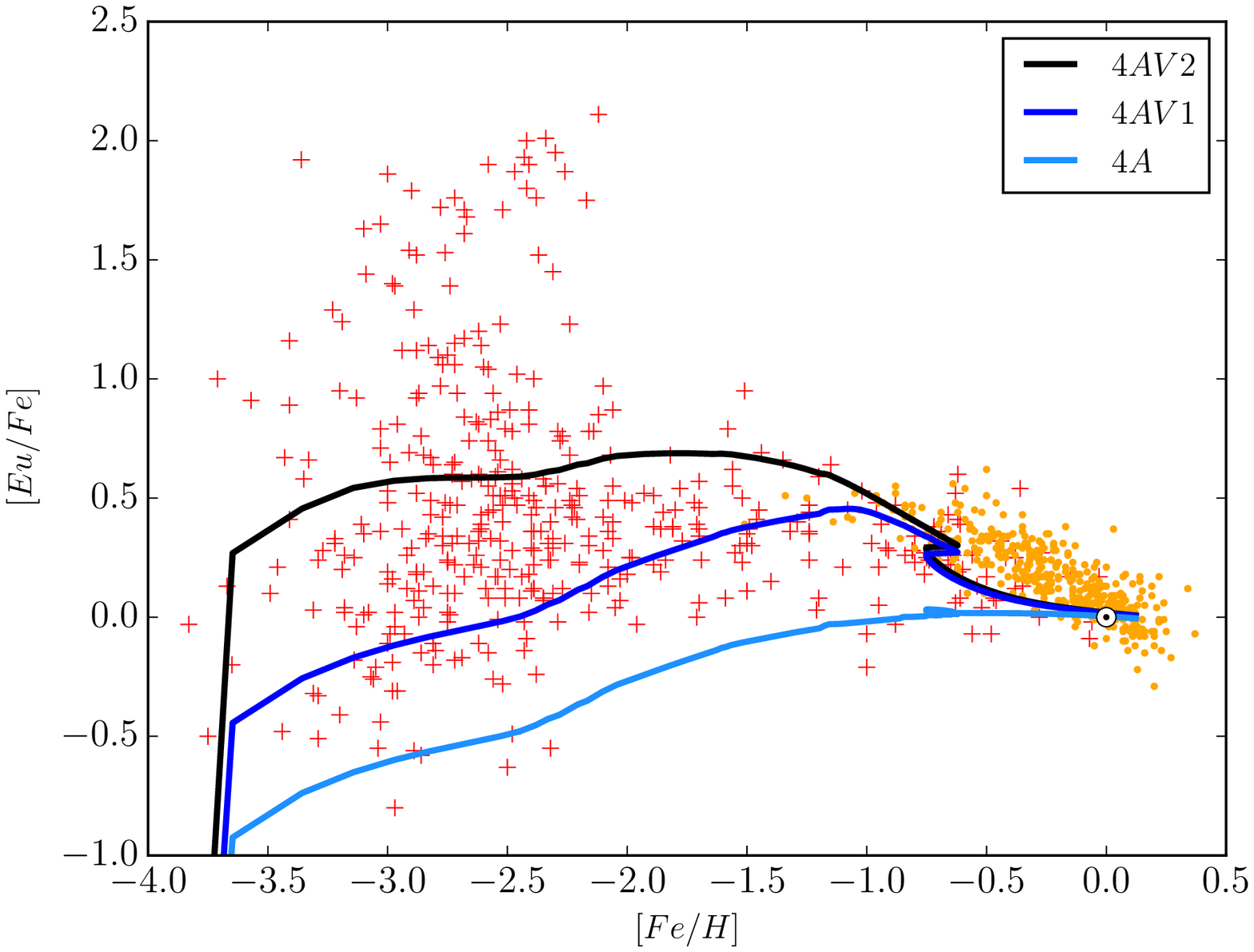}}
\subfloat[]{\includegraphics[width=0.5\textwidth]{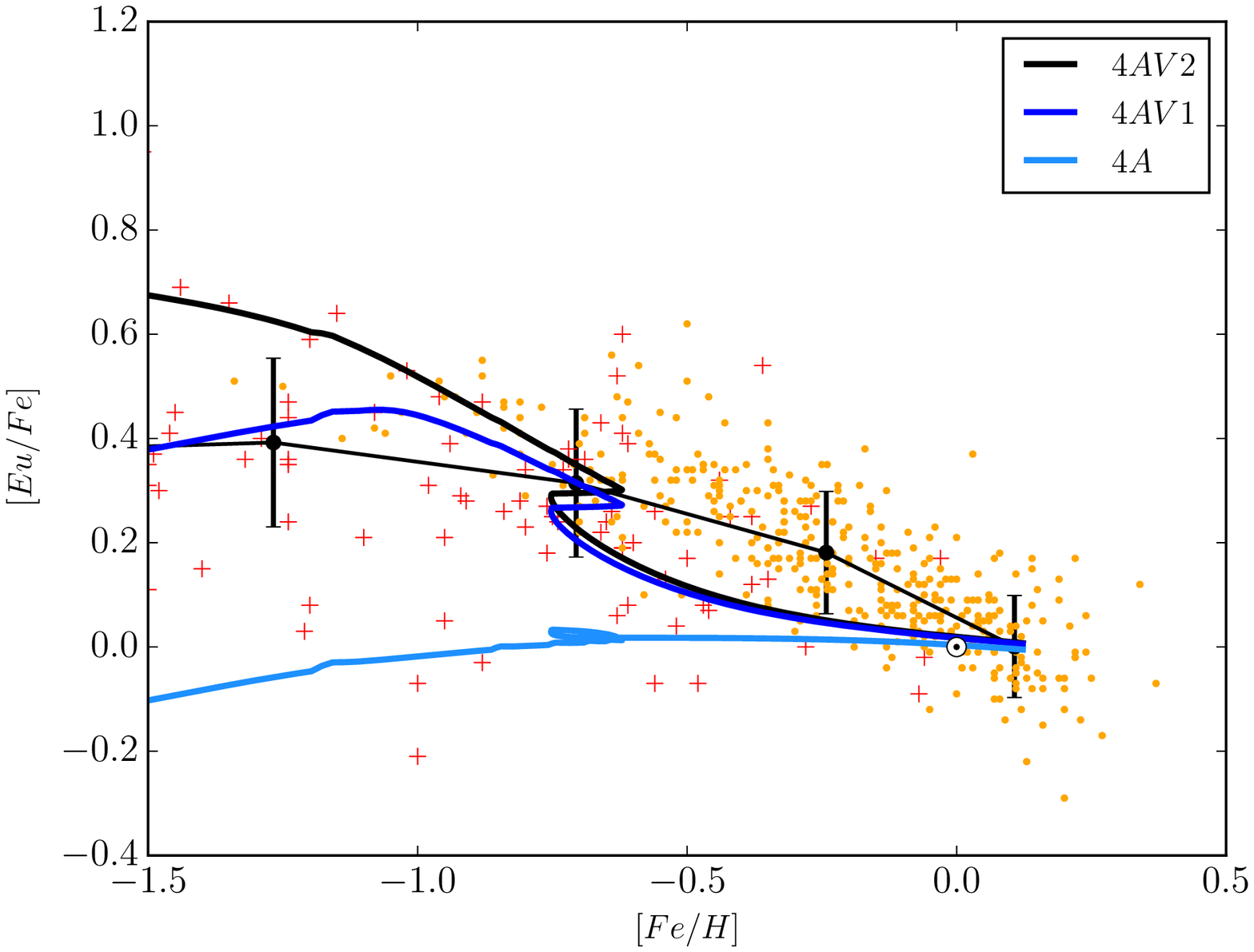}}
\caption{In these panels we show the effects of a variable MNS occurrence probability $\ams$. Models 4AV1 (blue line) and 4AV2 (black line) are variants of the model 4A (light blue line). In model 4AV1 we have adopted a step function for the $\ams$, while in model 4AV2 we have adopted a continuous function. Both of them make $\ams$ dependent on metallicity, traced by [Fe/H]. Panel b shows in detail the behavior of our models at [Fe/H]>-1.0.}
\label{sec5.evovar}
\end{figure*}

We have verified the ability to reproduce the current GW rate in the Milky Way and the absolute Eu abundance in the Sun. It happens that for both models 4AV1 and 4AV2, the predicted current MNS rate differs less than 0.1\% from the value reported in table \ref{cap5.models1}. Speaking about the absolute solar Eu abundance, the model 4AV1 predicts a value equal to $3.79\times 10^{-10}$ and the model 4AV2 predicts a value equal to $3.88\times 10^{-10}$. Therefore, we can conclude that the impact of this enhanced early occurrence on the current MNS rate and on the total amount of Eu in the Galaxy is negligible. On the other hand, a variable $\ams$ deeply influences the [Eu/Fe] vs [Fe/H] relation, as it is possible to see in figure \ref{sec5.evovar}. In fact, we are able to recover the decreasing trend in the disk stars, although the shape of the curve is not optimal (panel b). The model 4AV2 is also able to reproduce the low-metallicity plateau.

Finally, we have verified the impact of this new hypothesis on the cosmic rate of MNS. In order to do this, we have supposed two evolutionary patterns for the metallicity at high redshift. In the first one, the [Fe/H] evolves as predicted in the chemical evolution models of elliptical galaxies \citep{demasi18}. We have supposed a \citet{salpeter55} IMF, an exponential infall on timescales of 0.5 Gyr and a SFR efficiency of 20. These parameters produce a transition between primordial [Fe/H] values to [Fe/H]$\sim$-1.0 dex in 80 Myr. Instead in the second one, the [Fe/H] evolves as predicted for the Milky Way, taken as representative of the spiral galaxies, on a timescale four times larger. The real evolution of the cosmic metallicity is expected to be intermediate between these two extreme cases. The results are shown in figure \ref{sec5.SGRBvar}. The metallicity-dependent models differ from the non-metallicity dependent one only at very high redshift. Moreover, at least in case of a rapid chemical evolution, the predicted MNS rate does not differ too much neither at very early times. This fact demonstrates that a metallicity dependent $\ams$ does not invalidate our derived DTD with respect to the cosmic MNS rate.

\section{Conclusions}

In this paper, we have derived a new DTD for MNS starting from the initial separation and mass distributions of neutron star binary systems. We have assumed a flat distribution of NS masses and a power law distribution for the initial separations, with exponent $\beta$. We have tested four different values for $\beta$: 0.9, 0.0, -0.9 and -1.5, similar to what suggested for the DTD of SNeIa \citep{matteucci09} as derived from the double degenerate (DD) model of \citet{greggio05}.

We have tested our assumptions on the DTD of MNS and SNeIa on the following constraints:
\begin{enumerate}[i.]
\item the cosmic SGRB rate;
\item the evolution of Eu abundance in the Galaxy, in particular the [Eu/Fe] vs [Fe/H] relation.
\end{enumerate}
In this context, we have also tested the effect of core-collapse SNe as producers of Eu, besides MNS. Finally, we have tested a metallicity-dependent occurrence probability for MNS (the parameter $\ams$). Our main results can be summarized as follows:
\begin{enumerate}[i.]
\item Our derived DTD for MNS with $\beta$=-1.5 provides the best fit to the cosmic SGRB rate \citep[as found by][]{ghirlanda16}. Also our DTD with $\beta$=-0.9 and the DTD going like $\propto t^{-1}$ give a good fit. With these three DTDs, the average timescale of coalescence of MNS is 300-500 Myr. Shorter timescales produce too many events at high redshift (2-2.5), while longer timescales produce too few events over the entire range of redshift. This result is in agreement with the one found by Fong et al. (2017) that suggest a DTD $\propto t^{-1}$ for MNS to explain the proportion of SGRB incoming from early-type galaxies.
\item The other redshift distribution found by Zhang \& Wang (2017) can be reproduced only with a very top heavy (i.e with a lot of binaries with large separations) initial separation distribution, that corresponds to use our DTD with $\beta=6.0$. By the way, this extreme DTD cannot explain the Eu abundances in the Milky Way, independently from the contribution of CC-SNe to the Galactic Eu abundance.
\item When the MNS are the sole producers of Eu in the Galaxy, the models which better reproduce the [Eu/Fe] vs [Fe/H] pattern are those with short (< 30 Myr) coalescence timescales, in particular the constant total delay (stellar lifetime plus coalescence time) of 10 Myr and the $\propto t^{-2}$ distribution, in agreement with \citet{matteucci14} and \citet{cote18b}. On the other hand, ourderived  DTD is unable to reproduce the decreasing trend of [Eu/Fe] vs [Fe/H] observed in disk stars for reasonable values of the parameter $\beta$.
  
\item When the MNS are the sole producers of Eu in the Galaxy, their yield should be no more than $4.0 \times 10^{-6}$ $\ms$ per event. Higher yields overestimate the absolute Eu abundance in the Sun. This is in agreement with the lower end of the estimations of \citet{evans17} and \citet{troja17} in the kilonova AT2017gfo, but much higher than the estimate of \citet{smartt17}.
\item When CC-SNe co-produce Eu under the prescriptions of \citet{argast04} (their model SN2050), they become the main production site and dominate the [Eu/Fe] vs [Fe/H] relation. This can reconcile the short timescale required to explain the Eu abundance in the Milky Way with the longer timescale required to explain the SGRB redshift distribution. However, in this case,  the yield of Eu per MNS event is reduced to $1.5 \times 10^{-6}$ and falls (slightly) below the range of value found by \citet{evans17} and \citet{troja17}.
\item The influence of different DTDs for SNeIa (at least of those tested here) on [Eu/Fe] vs [Fe/H] relation seems mostly negligible, and it is of the order of $\sim 10 \%$.
\item The occurrence probability $\ams$ found for SGRBs produces a Galactic rate similar to that found by \citet{kalogera04} from pulsar luminosities (of the order of $1-2\times 10^{-2}$). It turns out $\ams \sim 10^{-2}$, interestingly close to the occurrence probability of SNeIa as determined form the cosmic rate of SNIa in the local universe \citep{cappellaro15}.
\item A metallicity-dependent occurrence probability, $\ams$, can increase the production of Eu at early times and at low [Fe/H] values ($\le -1.0$), enabling a DTD with average coalescence timescales of 300-500 Myr to reproduce both the SGRB redshift distribution and the [Eu/Fe] vs [Fe/H] relation in the Milky Way, when the MNS are the sole Eu producers.
\end{enumerate}

In general, allowing for the CC-SNe to form Eu helps reconciling the necessity of a short timescale in Eu production with the long timescales required for the SGRBs. However, this choice makes single massive stars the main Eu production site, by contributing no less than $\sim$60\% to the total Eu present in the Sun, and this fact is in contrast with the current understanding about heavy r-process elements nucleosynthesis. On the other hand, a time decreasing occurrence probability of MNS provides instead the correct production rate of Eu, without the contribution of CC-SNe, but this assumption is somewhat arbitrary and needs to be checked in the future by means of detailed population synthesis models.

\begin{landscape}
\begin{table}
\begin{center}
\begin{tabular}{llllccc|ccc}
\hline
Name & Eu from & SNeIa & MNS &  $\alpha_{Ia}$ & $\ams$ & MNS Eu yield & Current MNS & $X_{F\!e}$ & $X_{E\!u}$  \\
 & CC-SNe & DTD & DTD & ($\times 10^{-2}$) & ($\times 10^{-2}$) & ($\times 10^{-6} M_{\odot}$) & rate ($\text{Myr}^{-1}$) & ($\times 10^{-3}$) & ($\times 10^{-10}$)\\
\hline
1A & no & Prop to $t^{-1}$ & Our DTD $\beta=0.9$ & 4.77 & 0.66 & 4.0 & 76& 1.38& 1.68 \\
2A & no & Prop to $t^{-1}$ & Our DTD $\beta=0.0$ & 4.77 & 0.81 & 4.0 & 84& 1.38& 2.29 \\
3A & no & Prop to $t^{-1}$ & Our DTD $\beta=-0.9$ & 4.77 & 1.02 & 4.0 & 94& 1.38& 3.15 \\
4A & no & Prop to $t^{-1}$ & Our DTD $\beta=-1.5$ & 4.77 & 1.18 & 4.0 & 102& 1.38& 3.77 \\
5A & no & Prop to $t^{-1}$ & Prop to $t^{-1}$ & 4.77 & 1.09 & 4.0 & 93& 1.38& 3.27 \\
6A & no & Prop to $t^{-1}$ & Prop to $t^{-1.5}$ & 4.77 & 1.60 & 4.0 & 104& 1.38& 4.65 \\
7A & no & Prop to $t^{-1}$ & Prop to $t^{-2}$ & 4.77 & 1.73 & 4.0 & 93& 1.38& 4.28 \\
8A & no & Prop to $t^{-1}$ & Constant 10 Myr & 4.77 & 1.75 & 4.0 & 94& 1.38& 4.29 \\
1B & no & G05 $\beta_a=-0.9$ & Our DTD $\beta=0.9$ & 4.67 & 0.66 & 4.0 & 76& 1.52& 1.68 \\
2B & no & G05 $\beta_a=-0.9$ & Our DTD $\beta=0.0$ & 4.67 & 0.81 & 4.0 & 84& 1.52& 2.29 \\
3B & no & G05 $\beta_a=-0.9$ & Our DTD $\beta=-0.9$ & 4.67 & 1.02 & 4.0 & 94& 1.52& 3.15 \\
4B & no & G05 $\beta_a=-0.9$ & Our DTD $\beta=-1.5$ & 4.67 & 1.18 & 4.0 & 102& 1.52& 3.77 \\
5B & no & G05 $\beta_a=-0.9$ & Prop to $t^{-1}$ & 4.67 & 1.09 & 4.0 & 93& 1.52& 3.27 \\
6B & no & G05 $\beta_a=-0.9$ & Prop to $t^{-1.5}$ & 4.67 & 1.60 & 4.0 & 104& 1.52& 4.65 \\
7B & no & G05 $\beta_a=-0.9$ & Prop to $t^{-2}$ & 4.67 & 1.73 & 4.0 & 93& 1.52& 4.28 \\
8B & no & G05 $\beta_a=-0.9$ & Constant 10 Myr & 4.67 & 1.75 & 4.0 & 94& 1.52& 4.29 \\
\hline
1AS & yes & Prop to $t^{-1}$ & Our DTD $\beta=0.9$ & 4.77 & 0.66 & 1.5 & 76& 1.38& 2.57 \\
2AS & yes & Prop to $t^{-1}$ & Our DTD $\beta=0.0$ & 4.77 & 0.81 & 1.5 & 84& 1.38& 2.80 \\
3AS & yes & Prop to $t^{-1}$ & Our DTD $\beta=-0.9$ & 4.77 & 1.02 & 1.5 & 94& 1.38& 3.13 \\
4AS & yes & Prop to $t^{-1}$ & Our DTD $\beta=-1.5$ & 4.77 & 1.18 & 1.5 & 102& 1.38& 3.22 \\
5AS & yes & Prop to $t^{-1}$ & Prop to $t^{-1}$ & 4.77 & 1.09 & 1.5 & 93& 1.38& 3.11 \\
6AS & yes & Prop to $t^{-1}$ & Prop to $t^{-1.5}$ & 4.77 & 1.60 & 1.5 & 104& 1.38& 3.68 \\
7AS & yes & Prop to $t^{-1}$ & Prop to $t^{-2}$ & 4.77 & 1.73 & 1.5 & 93& 1.38& 3.55 \\
8AS & yes & Prop to $t^{-1}$ & Constant 10 Myr & 4.77 & 1.75 & 1.5 & 94& 1.38& 3.56 \\
1BS & yes & G05 $\beta_a=-0.9$ & Our DTD $\beta=0.9$ & 4.67 & 0.66 & 1.5 & 76& 1.52& 2.57 \\
2BS & yes & G05 $\beta_a=-0.9$ & Our DTD $\beta=0.0$ & 4.67 & 0.81 & 1.5 & 84& 1.52& 2.80 \\
3BS & yes & G05 $\beta_a=-0.9$ & Our DTD $\beta=-0.9$ & 4.67 & 1.02 & 1.5 & 94& 1.52& 3.13 \\
4BS & yes & G05 $\beta_a=-0.9$ & Our DTD $\beta=-1.5$ & 4.67 & 1.18 & 1.5 & 102& 1.52& 3.22 \\
5BS & yes & G05 $\beta_a=-0.9$ & Prop to $t^{-1}$ & 4.67 & 1.09 & 1.5 & 93& 1.52& 3.11 \\
6BS & yes & G05 $\beta_a=-0.9$ & Prop to $t^{-1.5}$ & 4.67 & 1.60 & 1.5 & 104& 1.52& 3.68 \\
7BS & yes & G05 $\beta_a=-0.9$ & Prop to $t^{-2}$ & 4.67 & 1.73 & 1.5 & 93& 1.52& 3.55 \\
8BS & yes & G05 $\beta_a=-0.9$ & Constant 10 Myr & 4.67 & 1.75 & 1.5 & 94& 1.52& 3.56 \\
\hline
\end{tabular}
\end{center}
\caption{The table reports all the models shown in the graphs. As explained in section 5, the occurrence probabilities for SNeIa ($\alpha_{Ia}$) has been tuned to obtain the current Galactic rate of SNeIa and the current cosmic rate MNS. Instead, the occurrence probability $\ams$ for MNS has been taken from table 1. The yield of Eu from MNS ($7^{th}$ column) has been tuned to reproduce the abundance in our Sun (the results of our simulations are reported in the $9^{th}$ and the $10^{th}$ columns). Finally, in the $8^{th}$ column is reported the predicted current rate of MNS in the Milky Way.}
\label{cap5.models1}
\end{table}
\end{landscape}

\section*{Acknowledgements}

PS acknowledges Carlo De Masi and Benoit C\^ot\'e for some useful suggestions. LG thanks Riccado Ciolfi for stimulating scientific discussions. FM acknowledges funds from University of Trieste (FRA2016). GC acknowledges financial support from the European Union Horizon 2020 research and innovation programme under the Marie Sk\l odowska-Curie grant agreement No 664931. This work has been partially supported by the the EU COST Action CA16117 (ChETEC).




\nocite{battistiniCat}
\bibliographystyle{mnras}
\bibliography{MNS_DTD} 



%
%
%
%

\bsp	
\label{lastpage}
\end{document}